\documentclass[oneside, reqno]{amsart}

\usepackage{xypic}
\input xy
\xyoption{all}
\usepackage{epsfig}
\usepackage{amsthm}
\usepackage{amssymb}
\usepackage{amsmath}
\usepackage{amscd}
\usepackage{color}
\usepackage[T1]{fontenc}
\usepackage{stackengine}
\usepackage{upgreek}
\usepackage[font=scriptsize]{caption}
\usepackage{wrapfig}
\stackMath

\usepackage{graphicx}
\usepackage{subfigure}

\newcommand{\bg}{\begin{equation}}
\newcommand{\ed}{\end{equation}}
\newcommand{\bga}{\begin{eqnarray}}
\newcommand{\eda}{\end{eqnarray}}
\newcommand{\pf}{\textbf{Proof:\ }}

\def\cbdu{\par{\raggedleft$\Box$\par}}

\newtheorem {Theorem}  {Theorem}

\numberwithin{Theorem}{section}

\newtheorem {Lemma}[Theorem]  {Lemma}

\theoremstyle{definition}
\newtheorem{Definition}[Theorem]{Definition}
\theoremstyle{remark}
\newtheorem{Remark}[Theorem]{\bf Remark}
\newtheorem{Conjecture}[Theorem]{\bf Conjecture}

\newtheorem {Corollary}[Theorem]{\bf Corollary}

\expandafter\chardef\csname pre amssym.def
at\endcsname=\the\catcode`\@ \catcode`\@=11
\def\undefine#1{\let#1\undefined}
\def\newsymbol#1#2#3#4#5{\let\next@\relax
 \ifnum#2=\@ne\let\next@\msafam@\else
 \ifnum#2=\tw@\let\next@\msbfam@\fi\fi
 \mathchardef#1="#3\next@#4#5}
\def\mathhexbox@#1#2#3{\relax
 \ifmmode\mathpalette{}{\m@th\mathchar"#1#2#3}%
 \else\leavevmode\hbox{$\m@th\mathchar"#1#2#3$}\fi}
\def\hexnumber@#1{\ifcase#1 0\or 1\or 2\or 3\or 4\or 5\or 6\or 7\or 8\or
 9\or A\or B\or C\or D\or E\or F\fi}

\font\teneufm=eufm10 \font\seveneufm=eufm7 \font\fiveeufm=eufm5
\newfam\eufmfam
\textfont\eufmfam=\teneufm \scriptfont\eufmfam=\seveneufm
\scriptscriptfont\eufmfam=\fiveeufm

\catcode`\@=\csname pre amssym.def at\endcsname

\newcounter{remark}
\def  \12  {{\frac{1}{2}}}

\def\build#1_#2^#3{\mathrel{\mathop{\kern 0pt#1}\limits_{#2}^{#3}}}

\begin{document}


\title[Dyadic MHD and NSE models]{Kolmogorov's dissipation number and determining wavenumber for dyadic models}


\author [Mimi Dai]{Mimi Dai}
\address{ School of Mathematics, Institute for Advanced Study, Princeton, NJ 08540, USA  \& 
Department of Mathematics, Statistics and Computer Science, University of Illinois at Chicago, Chicago, IL 60607, USA}
\email{mdai@uic.edu}

\author [Margaret Hoeller]{Margaret Hoeller}

\address{Department of Mathematics, Statistics and Computer Science, University of Illinois at Chicago, Chicago, IL 60607, USA}
\email{hoeller2@uic.edu} 

\author [Qirui Peng]{Qirui Peng}

\address{Department of Mathematics, Statistics and Computer Science, University of Illinois at Chicago, Chicago, IL 60607, USA}
\email{qpeng9@uic.edu} 

\author [Xiangxiong Zhasng]{Xiangxiong Zhang}

\address{Department of Mathematics, Purdue University, West Lafayette, IN 47907, USA}
\email{zhan1966@purdue.edu}






\begin{abstract}

We study some dyadic models for incompressible magnetohydrodynamics and Navier-Stokes equation. The existence of fixed point and stability of the fixed point are established. The scaling law of Kolmogorov's dissipation wavenumber arises from heuristic analysis. In addition, a time-dependent determining wavenumber is shown to exist; moreover, the time average of the determining wavenumber is proved to be bounded above by Kolmogorov's dissipation wavenumber. Additionally, based on the knowledge of the fixed point and stability of the fixed point, numerical simulations are performed to illustrate the energy spectrum in the inertial range below Kolmogorov's dissipation wavenumber.

\bigskip

KEY WORDS: dyadic models; intermittency; fixed point; dissipation wavenumber; determining wavenumber.

\hspace{0.02cm}CLASSIFICATION CODE: 35Q35, 76D03, 76W05.
\end{abstract}

\maketitle

\section{Introduction}

As a primary model in geophysics and astrophysics, the incompressible magnetohydrodynamics (MHD) system  
\begin{equation}\label{mhd}
\begin{split}
u_t+u\cdot\nabla u-B\cdot\nabla B+\nabla p=&\ \nu\Delta u,\\
B_t+u\cdot\nabla B-B\cdot\nabla u =&\ \mu\Delta B,\\
\nabla \cdot u= 0, \ \ \nabla \cdot B=&\ 0, 
\end{split}
\end{equation}
has been extensively studied both in physics and mathematics literatures. In the equations, the vector fields $u$ and $B$ denote respectively the fluid velocity and magnetic field; the scalar function $p$ represents the pressure; the parameters $\nu$ and $\mu$ stand for the kinematic viscosity and magnetic resistivity respectively. When $B\equiv0$, system (\ref{mhd}) reduces to the Navier-Stokes equation (NSE)
\begin{equation}\label{nse}
\begin{split}
u_t+u\cdot\nabla u+\nabla p=&\ \nu\Delta u,\\
\nabla \cdot u= &\ 0, 
\end{split}
\end{equation}
which has been a central objective in the area of fluid dynamics. 

Although being studied for a long time, there are still peculiar behaviour of solutions to (\ref{mhd}) and (\ref{nse}) which remain mysteries. The nonlinearities within the systems stir up complexities in the energy cascade mechanism of the dynamics. Many attempts to better understand the pure fluid system (\ref{nse}) involve studying approximating systems of (\ref{nse}). Notably, a class of dyadic models of (\ref{nse}) have been proposed and well studied, see 
\cite{ BMR, Bif, Ch, CF, CLT, DS, Fri, Gle, LPPPV, Ob, OY} (which is certainly not a complete list). Dyadic models for MHD have appeared and been studied mainly numerically in the physics literature, see \cite{Bis94, GLPG}, the survey paper \cite{PSF} and references therein. Recently, dyadic models for MHD were also proposed within the umbrella of harmonic analysis techniques in \cite{Dai-20} by the first author. The intermittency dimension $\delta$, introduced in term of the saturation of Bernstein's inequality \cite{CD-Kol}, is naturally included in the modeling process.



In this paper, we focus on two particular dyadic models suggested in \cite{Dai-20} for the MHD with external forcing, one with only forward energy cascade given by
\begin{equation}\label{sys-2}
\begin{split}
\frac{d}{dt}a_j+\nu\lambda_j^2 a_j=& \ \lambda_{j-1}^{\theta}a_{j-1}^2-\lambda_j^{\theta}a_ja_{j+1}
+\lambda_{j-1}^{\theta}b_{j-1}^2-\lambda_{j}^{\theta}b_jb_{j+1} +f_j,\\
\frac{d}{dt}b_j+\mu\lambda_j^2 b_j= & \ \lambda_j^{\theta}a_jb_{j+1}-\lambda_{j}^{\theta}b_ja_{j+1}
\end{split}
\end{equation}
for $j\geq 0$ and $a_{-1}=b_{-1}=0$; and another with both forward and backward energy cascades,
\begin{equation}\label{sys-3}
\begin{split}
\frac{d}{dt}a_j+\nu\lambda_j^2 a_j=& \ \lambda_{j-1}^{\theta}a_{j-1}^2-\lambda_j^{\theta}a_ja_{j+1}
-\lambda_{j-1}^{\theta}b_{j-1}^2+\lambda_{j}^{\theta}b_jb_{j+1} +f_j,\\
\frac{d}{dt}b_j+\mu\lambda_j^2 b_j= & -\lambda_j^{\theta}a_jb_{j+1}+\lambda_{j}^{\theta}b_ja_{j+1}.
\end{split}
\end{equation}
The parameter $\theta$ is given by
 \[\theta=\frac{2+n-\delta}{2}\in\left[1,\frac{2+n}2\right],  \ \ \ \delta\in[0,n]\]
with $\delta$ being the intermittency dimension of the $n$-dimensional velocity field $u$ and magnetic field $B$, which are assumed to be the same. The models (\ref{sys-2}) and (\ref{sys-3}) are equivalent to a class of MHD shell models proposed by physicists in \cite{GLPG}, which have been investigated subsequently in other work from physics community, including \cite{PSF}. Although these dyadic models suppress the spatial structure of MHD and do not preserve geometry features, their dynamics do reflect those of the original MHD turbulence. In particular, numerical studies in \cite{GLPG, PSF} capture intermittency statistics and chaotic behaviour which are consistent with experimental MHD turbulence.

If $b_j\equiv 0$ for all $j\geq 0$, both models (\ref{sys-2}) and (\ref{sys-3}) reduce to a forced dyadic model for the NSE (\ref{nse}),
\begin{equation}\label{nse-d}
\frac{d}{dt}a_j+\nu\lambda_j^2 a_j=\lambda_{j-1}^{\theta}a_{j-1}^2-\lambda_j^{\theta}a_ja_{j+1} +f_j,\ \ \ j\geq 0
\end{equation}
with $a_{-1}=0$, which is equipped with a forward energy cascade mechanism. 
For system (\ref{sys-2}) with $\nu=\mu=0$, the total energy defined by
\begin{equation}\notag
E(t)=\frac12\sum_{j\geq 1}\left(a_j^2(t)+b_j^2(t)\right)
\end{equation}
is formally conserved; however, the cross helicity defined by
\begin{equation}\notag
H^c(t)= \sum_{j\geq 1} a_j(t)b_j(t)
\end{equation}
is not conserved, accompanied with the mechanism of only forward energy cascade. Both the total energy and cross helicity are invariant for system (\ref{sys-3}) with $\nu=\mu=0$.

The dyadic NSE model (\ref{nse-d}) has been understood comprehensively. It was shown in \cite{Ch} that without external forcing: starting from nonnegative initial data, a strong solution exists globally in time when $\theta\leq 2$ and that solution blows up at finite time when $\theta>3$. Later the authors of \cite{BMR} proved the existence of global strong solution of (\ref{nse-d}) with zero forcing when $\theta\leq \frac52$. 
In \cite{CF}, assuming non-zero forcing only on the first mode, the authors proved the existence of a unique fixed point for (\ref{nse-d}); furthermore, they showed the fixed point is a global attractor and it converges to the global attractor of the inviscid system as $\nu\to 0$. In these findings, the preservation of positivity of solutions play vital roles. With an external forcing, uniqueness of Leray-Hopf solution was established for (\ref{nse-d}) with $\theta\leq 2$ in \cite{Fi, FK}; and non-unique Leray-Hopf solutions were constructed for (\ref{nse-d}) with $\theta>2$ in \cite{FK}.  We mention that, for the 2D NSE with certain forcing on the lowest mode, it was shown in \cite{Mar} that there is a globally attractive stationary steady state for any Reynolds number.

As a contrast, the dyadic MHD models (\ref{sys-2}) and (\ref{sys-3}) are much harder to be analyzed compared to the NSE model (\ref{nse-d}). Obviously, the interactions between the velocity components $a_j$ and magnetic field components $b_j$ create obstacles to understanding energy transfer from shell to shell. Moreover, due to such interactions, it is not clear whether a solution of (\ref{sys-2}) or (\ref{sys-3}) with positive initial data stays positive for all the time or not; a solution starting from positive initial data may become negative. Without the conservation of positivity of a solution, many techniques and methods for the dyadic NSE model fail to work for models (\ref{sys-2}) and (\ref{sys-3}). Nevertheless, with the lack of positivity, the authors of \cite{DF} proved existence and uniqueness of Leray-Hopf solution for (\ref{sys-2}) and (\ref{sys-3}) with $\theta\leq 2$ and appropriate forcing; they also constructed non-unique Leray-Hopf solutions for (\ref{sys-2}) and (\ref{sys-3}) with $\theta> 2$ and a particular forcing term. Assuming positivity, it was shown in \cite{Dai-21} that positive solution of (\ref{sys-2}) with large initial data blows up at finite time when $\theta>3$.

The main objective of this paper is to study large time behaviour of solutions to (\ref{sys-2}) and (\ref{sys-3}) in the following means. We first show the existence of a fixed point and study the stability of the fixed point under an assumption of either small forcing or small initial perturbation. In analogy with Kolmogorov's turbulence theory for hydrodynamics \cite{K41}, we then perform heuristic scaling analysis for the dyadic MHD models to indicate the existence of critical kinetic dissipation wavenumber $\kappa_{\mathrm d}^u$ and critical magnetic dissipation wavenumber $\kappa_{\mathrm d}^b$, with the former separating the kinetic inertial range from the kinetic dissipation range and the latter separating the ion-inertial range from the magnetic dissipation range. Below the critical wavenumber $\kappa_{\mathrm d}^u$ and $\kappa_{\mathrm d}^b$, we predict the scaling of the kinetic energy spectrum $\mathcal E_u(k)$ and magnetic energy spectrum $\mathcal E_b(k)$ respectively. 
Furthermore, we establish the existence of time-dependent determining wavenumber $\lambda_J$ in the sense: if two weak solutions in the energy space are close to each other below $\lambda_J$, the two solutions are asymptotically identical in the energy space. Both point-wise and time average estimates for the determining wavenumber $\lambda_J$ will be obtained. Notably, we show that the time average of $\lambda_J$ is bounded above by the dissipation wavenumber $\kappa_{\mathrm d}^u$ and $\kappa_{\mathrm d}^b$. It thus provides theoretical ground for the heuristic scaling law of $\kappa_{\mathrm d}^u$ and $\kappa_{\mathrm d}^b$. In the end, based on the knowledge of the fixed point and stability results established earlier, we design a numerical scheme to simulate the dyadic models. The numerical results confirm the predicted scaling law of the energy spectra $\mathcal E_u(k)$ and $\mathcal E_b(k)$.

We note that the dyadic NSE model (\ref{nse-d}) is a reduced system of both (\ref{sys-2}) and (\ref{sys-3}). Thus the results described in the last paragraph for the dyadic MHD models can be obtained for (\ref{nse-d}) with some modifications.


\bigskip

\section{Preliminaries}

We denote $H=l^2$ which is endowed with the standard scalar product and norm,
\[(u,v):=\sum_{n=1}^\infty u_nv_n, \ \ \ |u|:=\sqrt{(u,u)}.\]
As mentioned earlier, we choose the wavenumber $\lambda_n=\lambda^n/L$ for a constant $\lambda>1$ and all integers $n\geq 0$, where $L$ denotes the domain size for the original NSE and MHD systems.
As standard Sobolev space $H^s$ for functions with spatial variables, we
use the same notation $H^s$ in this paper to represent the space for a sequence $\{u_n\}_{n=1}^{\infty}$, which is endowed with the scaler product 
\[(u,v)_s:=\sum_{n=1}^\infty \lambda_n^{2s}u_nv_n\]
and the norm
\[\|u\|_s:=\sqrt{(u,u)_s}.\]
Notice that $H^0=l^2$ is the energy space. 

We introduce the concept of solutions for dyadic models as follows. 
\begin{Definition}\label{def1}
A pair of $H$-valued functions $(a(t), b(t))$ defined on $[t_0,\infty)$ is said to be a weak solution of (\ref{sys-2}) if $a_j$ and $b_j$ satisfy (\ref{sys-2}) and $a_j, b_j\in C^1([t_0,\infty))$ for all $j\geq0$.
\end{Definition}

\begin{Definition}\label{def2}
A solution $(a(t), b(t))$ of (\ref{sys-2}) is strong on $[T_1, T_2]$ if $\|a\|_1$ and $\|b\|_1$ are bounded on $[T_1, T_2]$. A solution is strong on $[T_1, \infty)$ if it is strong on every interval $[T_1, T_2]$ for any $T_2>T_1$.
\end{Definition}

\begin{Definition}\label{def3}
A Leray-Hopf solution $(a(t), b(t))$ of (\ref{sys-2}) on $[t_0,\infty)$ is a weak solution satisfying the energy inequality
\begin{equation}\notag
|a(t)|^2+|b(t)|^2+2\nu\int_{t_1}^t\|a(\tau)\|_1 \, d\tau+2\mu\int_{t_1}^t\|b(\tau)\|_1 \, d\tau\leq |a(t_1)|^2+|b(t_1)|^2
\end{equation}
for all $t_0\leq t_1\leq t$ and a.e. $t_1\in[t_0,\infty)$.
\end{Definition}

Solutions of (\ref{sys-3}) and (\ref{nse-d}) are defined in the same way.  

\begin{Definition}\label{def-stab1} (Lyapunov stability)
The system 
\begin{equation}\notag
x'(t)=f(x), \ \ x(t_0)=x_0
\end{equation}
is said to be stable in the sense of Lyapunov with respect to the fixed point $x^*$, if for any $\varepsilon>0$, there exists a constant $\delta=\delta(\varepsilon, t_0)>0$ such that 
\begin{equation}\notag
|x(t)-x^*|<\varepsilon, \ \ \ \forall \ \ t\geq t_0  \ \ \mbox{provided} \ \ |x(t_0)-x^*|< \delta.
\end{equation}
\end{Definition}

\bigskip

\section{Fixed point}
\label{sec-fix}

\subsection{Existence of fixed point and properties of fixed point}

We consider the stationary system of (\ref{sys-2}) with 
\[\frac{d}{dt}a_j=\frac{d}{dt}b_j=0, \ \ \forall j\geq0,\]
\[f_0>0; \ \ f_j=0, \ \forall j\geq1\] 
Define the rescaled quantities
\[A_j=\lambda^{-\frac16\theta}f_0^{-\frac12}\lambda_j^{\frac13\theta}a_j, \ \ \ B_j=\lambda^{-\frac16\theta}f_0^{-\frac12}\lambda_j^{\frac13\theta}b_j,\]
\[\bar\nu=\nu\lambda^{\frac16\theta}f_0^{-\frac12}, \ \ \ \bar\mu=\mu\lambda^{\frac16\theta}f_0^{-\frac12}.\]
The stationary system of (\ref{sys-2}) can be written as 
\begin{equation}\label{stat-2-gen}
\begin{split}
A_{j-1}^2-A_jA_{j+1}+B_{j-1}^2-B_jB_{j+1}=&\ \bar\nu \lambda_j^{2-\frac23\theta}A_j, \ \ j\geq 1,\\
A_jB_{j+1}-B_jA_{j+1}
=&\ \bar\mu \lambda_j^{2-\frac23\theta}B_j, \ \ j\geq 1,\\
-A_0A_1-B_0B_1=&\ \bar\nu A_0-1,\\
A_0B_1-B_0A_1=&\ \bar\mu B_0. 
\end{split}
\end{equation}

\begin{Theorem}\label{thm-ex-2}
There exists a fixed point $(a^*,b^*)\in \ell^2\times \ell^2$ to (\ref{sys-2}).
Equivalently, there exists a solution $\{(A_j,B_j)\}$ to (\ref{stat-2-gen}) with $(A,B)\in H^{-\frac13\theta}\times H^{-\frac13\theta}$.  
\end{Theorem}
\pf
It is sufficient to show the existence of an absorbing ball for (\ref{sys-2}). Consequently, the existence of solutions to the stationary system follows from standard fixed point arguments, for instance, an application of Brouwer's fixed point theorem. 

To make the following argument rigorous, one can apply it to Galerkin approximating systems and pass to the limit. Formally, multiplying $a_j$ to the first equation of (\ref{sys-2}), $b_j$ to the second one, adding the resulted equations, and taking sum over $j\geq 1$ gives us the {\it a priori} energy estimate, 
\begin{equation}\notag
\begin{split}
\frac{d}{dt}\left(\|a\|_2^2+\|b\|_2^2\right)\leq&\ -2\nu\sum_{j=0}^\infty \lambda_j^2a_j^2-2\mu\sum_{j=0}^\infty \lambda_j^2b_j^2+\sum_{j=0}^\infty a_jf_j\\
\leq&\ -2m_0 \lambda_0^2\left(\|a\|_2^2+\|b\|_2^2\right)+m_0\lambda_0^2\|a\|_2^2+(m_0\lambda_0^2)^{-1} \|f\|_2^2\\
\leq&\ -m_0 \lambda_0^2\left(\|a\|_2^2+\|b\|_2^2\right)+(m_0\lambda_0^2)^{-1}\|f\|_2^2
\end{split}
\end{equation}
where $m_0=\min\{\nu,\mu\}$. 
It follows from the energy estimate that
\[\frac{d}{dt}\left(\|a\|_2^2+\|b\|_2^2\right)<0, \ \ \mbox{if}\ \ \|a\|_2^2+\|b\|_2^2>(m_0\lambda_0^2)^{-2}\|f\|_2^2.\]
Thus, there exists an absorbing ball in the energy space $\ell^2\times \ell^2$ with radius \\
$(m_0\lambda_0^2)^{-1}\|f\|_2$.

\cbdu

\begin{Remark}\label{rk-high-reg}

We note $(a^*, b^*)$ is a Leray-Hopf solution.
By Leray structure theory, we know every Leray-Hopf solution is regular almost everywhere in time. Since $(a^*, b^*)$ is stationary, it is thus regular everywhere and hence belongs to $H^s\times H^s$ for any $s>0$. Consequently, the solution $(A,B)$ of (\ref{stat-2-gen}) is in fact in $H^{s}\times H^{s}$ for any $s>0$ as well. 

\end{Remark}


\begin{Lemma}\label{le-positive}
Let $(A,B)$ be a solution to (\ref{stat-2-gen}). Then we have either
\[A_j>0,\ \  B_j>0, \ \ \ \forall \ \ j\geq 0,\]
or
\[A_j>0,\ \  B_j<0, \ \ \ \forall \ \ j\geq 0.\]
\end{Lemma}
\pf
Denote $(\ref{stat-2-gen})_i$ by the $i$-th equation of system (\ref{stat-2-gen}) for $1\leq i\leq 4$. Multiplying equation $(\ref{stat-2-gen})_1$ by $A_j$, $(\ref{stat-2-gen})_2$ by $B_j$, $(\ref{stat-2-gen})_3$ by $A_0$ and $(\ref{stat-2-gen})_4$ by $B_0$, and taking sum for $j\geq0$, we obtain
\begin{equation}\notag
\sum_{j=0}^\infty\lambda_j^{2-\frac23\theta}\left(\bar\nu A_j^2+\bar\mu B_j^2\right)=A_0.
\end{equation}
Since $(A,B)\in H^s\times H^s$ for any $s>0$, the sum on the left hand side of the equation above is well-defined.
If instead taking sum for $j\geq J$ with $J\geq 0$, we have
\begin{equation}\notag
\sum_{j=J}^\infty\lambda_j^{2-\frac23\theta}\left(\bar\nu A_j^2+\bar\mu B_j^2\right)=\left(A_{J-1}^2+B_{J-1}^2\right)A_J.
\end{equation}
Thus we have $A_j>0$ for all $j\geq 0$. 

The equations $(\ref{stat-2-gen})_2$ and $(\ref{stat-2-gen})_4$ imply
\begin{equation}\notag
\begin{split}
B_{j+1}=&\ \frac{\bar\mu \lambda_j^{2-\frac23\theta}+A_{j+1}}{A_j} B_j, \ \ \forall \ \ j\geq 1,\\
B_1=&\ \frac{\bar\mu+A_{1}}{A_0} B_0. 
\end{split}
\end{equation}
Since $A_j>0$ for all $j\geq 0$, we have if $B_0>0$, then $B_j> 0$ for all $j\geq 0$; or if $B_0<0$, then $B_j< 0$ for all $j\geq 0$.

\cbdu

\medskip

In the case of the inviscid non-resistive dyadic MHD model, i.e. $\nu=\mu=0$ in (\ref{sys-2}), we are able to derive an exact form of the fixed point, which will be used to provide the last mode for the truncated system in our numerical simulation. The stationary system (\ref{stat-2-gen}) becomes
\begin{equation}\label{stat-2-mhd}
\begin{split}
A_{j-1}^2-A_jA_{j+1}+B_{j-1}^2-B_jB_{j+1}=&\ 0, \ \ j\geq 1,\\
A_jB_{j+1}-B_jA_{j+1} =&\ 0, \ \ j\geq 1,\\
-A_0A_1-B_0B_1=& -1,\\
A_0B_1-B_0A_1=&\ 0.
\end{split}
\end{equation}
\begin{Lemma}\label{le-ex2} \cite{DF}
The solutions $\{(A_j,B_j)\}$ of (\ref{stat-2-mhd}) satisfy
\[A_j=A_0, \ \ B_j=B_0, \ \ \forall \ \ j\geq1,\]
\[A_0^2+B_0^2=1.\]
\end{Lemma}

\medskip

\subsection{Stability of fixed point}\label{sec-stability}
In this part, we show that a fixed point of (\ref{sys-2}) is stable under certain conditions. Specifically, we will establish the following results: (i) when the force $f_0$ is small or $\mu$ and $\nu$ are large enough, the fixed point is exponentially stable and thus unique; (ii) in general, i.e., without the smallness assumptions, solutions starting from initial data that is closed to a fixed point converge to the fixed point, that is, the fixed point is globally attracting in the sense of Lyapunov.

\begin{Theorem}\label{thm-stab1}
Solutions of system (\ref{sys-2}) with $\theta\leq 2$ and $0<f_0<\frac18 m_0^2$ converge to a fixed point in $\ell^2\times \ell^2$.
Consequently, the fixed point is unique.
\end{Theorem}
\pf
Let $(a(t), b(t))$ be a solution of (\ref{sys-2}). Denote the difference with the fixed point by
\[u(t)=a(t)-a^*, \ \ \ v(t)=b(t)-b^*.\]
The difference $(u,v)$ satisfies the system
\begin{equation}\label{eq-diff}
\begin{split}
\frac{d}{dt}u_j=&\ -\nu\lambda_j^2 u_j-\lambda_j^\theta u_ju_{j+1}+\lambda_{j-1}^\theta u_{j-1}^2
-\lambda_j^\theta v_jv_{j+1}+\lambda_{j-1}^\theta v_{j-1}^2\\
&-\lambda_j^\theta u_j a_{j+1}^*-\lambda_j^\theta a_j^* u_{j+1}+2\lambda_{j-1}^\theta u_{j-1} a_{j-1}^*\\
&-\lambda_j^\theta v_jb_{j+1}^*-\lambda_j^\theta b_j^* v_{j+1}+2\lambda_{j-1}^\theta v_{j-1} b_{j-1}^*,\\
\frac{d}{dt}v_j=&\ -\mu\lambda_j^2 v_j+\lambda_j^\theta u_j v_{j+1}-\lambda_j^\theta v_ju_{j+1}\\
&+\lambda_j^\theta u_j b_{j+1}^*+\lambda_j^\theta a_j^* v_{j+1}-\lambda_j^\theta v_ja_{j+1}^*
-\lambda_j^\theta b_j^* u_{j+1}.
\end{split}
\end{equation}
Multiplying $(\ref{eq-diff})_1$ by $u_j$ and $(\ref{eq-diff})_2$ by $v_j$, taking sum for $0\leq j\leq k$ with arbitrary $k\geq 1$, we get
\begin{equation}\notag
\begin{split}
\frac12\frac{d}{dt}\sum_{j=0}^k\left(u_j^2+v_j^2\right)=&\ -\sum_{j=0}^k \left(\nu \lambda_j^2u_j^2+\mu\lambda_j^2v_j^2\right)-\sum_{j=0}^k \lambda_j^\theta\left(u_j^2+v_j^2\right)a_{j+1}^*\\
&+2\sum_{j=0}^k\lambda_{j-1}^\theta u_{j-1}u_j a_{j-1}^*-\sum_{j=0}^k\lambda_{j}^\theta u_{j}u_{j+1} a_{j}^*\\
&+2\sum_{j=0}^k\lambda_{j-1}^\theta v_{j-1} u_j b_{j-1}^*-\sum_{j=0}^k\lambda_{j}^\theta v_{j} u_{j+1} b_{j}^*\\
&-\sum_{j=0}^k \lambda_j^\theta u_jv_{j+1} b_j^*+\sum_{j=0}^k\lambda_j^\theta v_jv_{j+1}a_j^*\\
&-\lambda_k^\theta u_k^2u_{k+1}-\lambda_k^\theta v_k^2u_{k+1}.
\end{split}
\end{equation}
Since $(a^*, b^*)\in H^s\times H^s$ for any $s>0$, $\lambda_j^sa_j^*\to 0$ and $\lambda_j^sb_j^*\to 0$ as $j\to \infty$. We also note that $\sum_{j=0}^\infty \lambda_j^2u_j^2$ and $\sum_{j=0}^\infty \lambda_j^2v_j^2$ are integrable. Hence, taking $k\to\infty$ in the equation above and applying the dominated convergence theorem, we deduce
\begin{equation}\label{est-energy-diff}
\begin{split}
&\frac12 \sum_{j=0}^\infty\left(u_j^2(t)+v_j^2(t)\right)-\frac12 \sum_{j=0}^\infty\left(u_j^2(0)+v_j^2(0)\right)\\
& + \int_0^t \sum_{j=0}^\infty \left(\nu \lambda_j^2u_j^2(\tau)+\mu\lambda_j^2v_j^2(\tau)\right)\, d\tau\\
\leq & - \int_0^t \sum_{j=0}^\infty \lambda_j^\theta\left(u_j^2(\tau)+v_j^2(\tau)\right)a_{j+1}^*\, d\tau\\
& +\int_0^t \sum_{j=0}^\infty \lambda_{j}^\theta u_{j}(\tau)u_{j+1}(\tau) a_{j}^*\, d\tau
 +\int_0^t \sum_{j=0}^\infty \lambda_{j}^\theta v_{j}(\tau)u_{j+1}(\tau) b_{j}^*\, d\tau\\
&-\int_0^t\sum_{j=0}^\infty \lambda_j^\theta u_j(\tau) v_{j+1}(\tau) b_j^*\, d\tau
+\int_0^t\sum_{j=0}^\infty \lambda_j^\theta v_j(\tau) v_{j+1}(\tau) a_j^*\, d\tau.
\end{split}
\end{equation}
From the proof of Theorem \ref{thm-ex-2}, the fixed point lies in an absorbing ball with radius $m_0^{-1} \lambda_0^{-2} \|f\|_2=(\min\{\nu,\mu\})^{-1} f_0$ in $\ell^2\times \ell^2$. It follows that 
\[\|a^*\|_2^2+\|b^*\|_2^2\leq \min\{\nu,\mu\}^{-2} f_0^2,\]
and hence 
\begin{equation}\label{est-fixed-2}
|a_j^*|\leq (\min\{\nu,\mu\})^{-1} f_0, \ \ \ |b_j^*|\leq (\min\{\nu,\mu\})^{-1} f_0, \ \ \ \forall \ \ j\geq 0.
\end{equation}
Applying (\ref{est-fixed-2}), we have
\begin{equation}\label{est-energy-diff-1}
\begin{split}
- \int_0^t \sum_{j=0}^\infty \lambda_j^\theta\left(u_j^2(\tau)+v_j^2(\tau)\right)a_{j+1}^*\, d\tau
\leq m_0^{-1}f_0  \int_0^t \sum_{j=0}^\infty \lambda_j^\theta\left(u_j^2(\tau)+v_j^2(\tau)\right)\, d\tau,
\end{split}
\end{equation}
\begin{equation}\label{est-energy-diff-3}
\begin{split}
\int_0^t \sum_{j=0}^\infty \lambda_{j}^\theta u_{j}(\tau)u_{j+1}(\tau) a_{j}^*\, d\tau
\leq 
m_0^{-1}f_0 \int_0^t \sum_{j=0}^\infty \lambda_j^{\theta}u_j^2(\tau)\, d\tau,
\end{split}
\end{equation}
\begin{equation}\label{est-energy-diff-4}
\begin{split}
\int_0^t \sum_{j=0}^\infty \lambda_{j}^\theta v_{j}(\tau)u_{j+1}(\tau) b_{j}^*\, d\tau
\leq m_0^{-1}f_0 \int_0^t \sum_{j=0}^\infty \lambda_j^{\theta}\left(u_j^2(\tau)+v_j^2(\tau)\right)\, d\tau,
\end{split}
\end{equation}
\begin{equation}\label{est-energy-diff-6}
\begin{split}
-\int_0^t \sum_{j=0}^\infty \lambda_{j}^\theta u_{j}(\tau)v_{j+1}(\tau) b_{j}^*\, d\tau
\leq m_0^{-1}f_0 \int_0^t \sum_{j=0}^\infty \lambda_j^{\theta}\left(u_j^2(\tau)+v_j^2(\tau)\right)\, d\tau,
\end{split}
\end{equation}
\begin{equation}\label{est-energy-diff-7}
\begin{split}
\int_0^t \sum_{j=0}^\infty \lambda_{j}^\theta v_{j}(\tau)v_{j+1}(\tau) a_{j}^*\, d\tau
\leq 
m_0^{-1}f_0 \int_0^t \sum_{j=0}^\infty \lambda_j^{\theta}v_j^2(\tau)\, d\tau.
\end{split}
\end{equation}
Combining (\ref{est-energy-diff}) and (\ref{est-energy-diff-1})-(\ref{est-energy-diff-7}), we infer
\begin{equation}\label{est-energy-diff-8}
\begin{split}
& \sum_{j=0}^\infty\left(u_j^2(t)+v_j^2(t)\right)- \sum_{j=0}^\infty\left(u_j^2(0)+v_j^2(0)\right)\\
\leq &\ - 2m_0\int_0^t \sum_{j=0}^\infty \lambda_j^2 \left(u_j^2(\tau)+v_j^2(\tau)\right)\, d\tau\\
&+8m_0^{-1}f_0 \int_0^t \sum_{j=0}^\infty \lambda_j^\theta \left(u_j^2(\tau)+v_j^2(\tau)\right)\, d\tau.
\end{split}
\end{equation}
If $\theta\leq 2$ and $0<f_0<\frac18 m_0^2$, inequality (\ref{est-energy-diff-8}) implies
\begin{equation}\label{est-energy-diff-9}
\begin{split}
& \sum_{j=0}^\infty\left(u_j^2(t)+v_j^2(t)\right)- \sum_{j=0}^\infty\left(u_j^2(0)+v_j^2(0)\right)\\
\leq &\ - m_0\int_0^t \sum_{j=0}^\infty \lambda_j^2 \left(u_j^2(\tau)+v_j^2(\tau)\right)\, d\tau\\
\leq &\ - m_0\int_0^t \sum_{j=0}^\infty \left(u_j^2(\tau)+v_j^2(\tau)\right)\, d\tau. 
\end{split}
\end{equation}
Employing Gr\"onwall's integral inequality to (\ref{est-energy-diff-9}), we obtain
\begin{equation}\notag
\sum_{j=0}^\infty\left(u_j^2(t)+v_j^2(t)\right)\leq e^{-m_0t} \sum_{j=0}^\infty\left(u_j^2(0)+v_j^2(0)\right).\end{equation}
It indicates that 
\[(u(t), v(t)) \to 0 \ \ \ \mbox{as} \ \ t\to\infty \ \ \mbox{in} \ \ \ell^2\times \ell^2,\]
and hence 
\[(a(t), b(t)) \to (a^*, b^*) \ \ \ \mbox{as} \ \ t\to\infty \ \ \mbox{in} \ \ \ell^2\times \ell^2.\]

\cbdu


\medskip

\begin{Theorem}\label{thm-stab2}
(Lyapunov Stability)
Let $(a(t),b(t))$ be a solution to (\ref{sys-2}) with initial data $(a_0, b_0)$ on $[0,T]$. 
For any $\varepsilon>0$, there exists a constant $\delta>0$ such that 
\[\|a(t)-a^*\|_2^2+\|b(t)-b^*\|_2^2\leq \varepsilon, \ \ \ \forall \ \ 0\leq t\leq T,\]
provided \[\|a_0-a^*\|_2^2+\|b_0-b^*\|_2^2\leq \delta.\]
\end{Theorem}
\pf
A slight modification of the proof of Theorem \ref{thm-stab1} is sufficient to justify the statement.
 We start from the energy estimate (\ref{est-energy-diff}) established earlier. Instead of (\ref{est-fixed-2}), applying the fact that $\lambda_j^sa_j^*\to 0$ and $\lambda_j^sb_j^*\to 0$ as $j\to \infty$ for any $s>0$ we know there exists $J>0$ such that 
\begin{equation}\label{Hs-decay}
\lambda_j^{\theta-2}a^*_j<\frac{m_0}{36}, \ \ \lambda_j^{\theta-2}b^*_j<\frac{m_0}{36}, \ \ \forall \ \ j> J.
\end{equation}
Denote 
\[C_1=8\lambda_J^{\theta}\sup_{0\leq j\leq J+1} \left(\|a^*_j\|_{L^\infty}+\|b^*_j\|_{L^\infty}\right).\]
Employing (\ref{Hs-decay}), we estimate the high modes part of the right hand side of (\ref{est-energy-diff}) as follows
\begin{equation}\notag
\begin{split}
- \int_0^t \sum_{j=J+1}^\infty \lambda_j^\theta\left(u_j^2(\tau)+v_j^2(\tau)\right)a_{j+1}^*\, d\tau
\leq \frac{m_0}{36} \int_0^t \sum_{j=J+1}^\infty \lambda_j^2\left(u_j^2(\tau)+v_j^2(\tau)\right)\, d\tau,
\end{split}
\end{equation}
\begin{equation}\notag
\begin{split}
\int_0^t \sum_{j=J+1}^\infty \lambda_{j}^\theta u_{j}(\tau)u_{j+1}(\tau) a_{j}^*\, d\tau
\leq 
\frac{m_0}{36} \int_0^t \sum_{j=J+1}^\infty \lambda_j^2u_j^2(\tau)\, d\tau,
\end{split}
\end{equation}
\begin{equation}\notag
\begin{split}
\int_0^t \sum_{j=J+1}^\infty \lambda_{j}^\theta v_{j}(\tau)u_{j+1}(\tau) b_{j}^*\, d\tau
\leq \frac{m_0}{36}  \int_0^t \sum_{j=J+1}^\infty \lambda_j^2\left(u_j^2(\tau)+v_j^2(\tau)\right)\, d\tau,
\end{split}
\end{equation}
\begin{equation}\notag
\begin{split}
-\int_0^t \sum_{j=J+1}^\infty \lambda_{j}^\theta u_{j}(\tau)v_{j+1}(\tau) b_{j}^*\, d\tau
\leq \frac{m_0}{36}  \int_0^t \sum_{j=J+1}^\infty \lambda_j^2\left(u_j^2(\tau)+v_j^2(\tau)\right)\, d\tau,
\end{split}
\end{equation}
\begin{equation}\notag
\begin{split}
\int_0^t \sum_{j=J+1}^\infty \lambda_{j}^{\theta} v_{j}(\tau)v_{j+1}(\tau) a_{j}^*\, d\tau
\leq 
\frac{m_0}{36}  \int_0^t \sum_{j=J+1}^\infty \lambda_j^2v_j^2(\tau)\, d\tau.
\end{split}
\end{equation}
While the low modes part of the right hand side of (\ref{est-energy-diff}) satisfies 
\begin{equation}\notag
\begin{split}
& - \int_0^t \sum_{j=0}^J \lambda_j^\theta\left(u_j^2(\tau)+v_j^2(\tau)\right)a_{j+1}^*\, d\tau\\
& +\int_0^t \sum_{j=0}^J \lambda_{j}^\theta u_{j}(\tau)u_{j+1}(\tau) a_{j}^*\, d\tau
 +\int_0^t \sum_{j=0}^J \lambda_{j}^\theta v_{j}(\tau)u_{j+1}(\tau) b_{j}^*\, d\tau\\
&-\int_0^t\sum_{j=0}^J \lambda_j^\theta u_j(\tau) v_{j+1}(\tau) b_j^*\, d\tau
+\int_0^t\sum_{j=0}^J \lambda_j^\theta v_j(\tau) v_{j+1}(\tau) a_j^*\, d\tau\\
\leq& \ C_1  \int_0^t \sum_{j=0}^J\left(u_j^2(\tau)+v_j^2(\tau)\right)\, d\tau.
\end{split}
\end{equation}
As a consequence, it follows from (\ref{est-energy-diff}) that
\begin{equation}\label{est-energy-diff-10}
\begin{split}
&\frac12 \sum_{j=0}^\infty\left(u_j^2(t)+v_j^2(t)\right)-\frac12 \sum_{j=0}^\infty\left(u_j^2(0)+v_j^2(0)\right)\\
\leq &\ - \frac12m_0\int_0^t \sum_{j=0}^\infty \left(\lambda_j^2u_j^2(\tau)+\lambda_j^2v_j^2(\tau)\right)\, d\tau
+C_1  \int_0^t \sum_{j=0}^J\left(u_j^2(\tau)+v_j^2(\tau)\right)\, d\tau.
\end{split}
\end{equation}
Applying Gr\"onwall's inequality to (\ref{est-energy-diff-10}) we obtain
\begin{equation}\notag
\sum_{j=0}^\infty\left(u_j^2(t)+v_j^2(t)\right)\leq e^{2C_1t}\sum_{j=0}^\infty\left(u_j^2(0)+v_j^2(0)\right). 
\end{equation}
The statement of the theorem follows immediately.

\cbdu





\bigskip

\section{Analogue of Kolmogorov's turbulence theory}
\label{sec-turbulence}


Denote $\varepsilon=\nu\langle\|\nabla u\|_{L^2}^2\rangle$ by the average energy dissipation rate per unit mass. The zeroth law of turbulence on anomalous energy dissipation postulates that $\varepsilon$ does not converge to 0 as $\nu\to 0$.
On this ground, Kolmogorov \cite{K41} proposed a theory for homogeneous isotropic turbulence under the assumption of self-similarity of the flow. The two major predictions are stated below.

\begin{Conjecture}(Kolmogorov's predictions)\\
(i)  There exists a critical wavenumber $\kappa_{\mathrm d}\sim (\varepsilon \nu^{-3})^{\frac14}$ for the 3D flow such that only the low frequency part below $\kappa_{\mathrm d}$ is essential to describe the flow.\\
(ii) The energy density spectrum has the form
\begin{equation}\notag
\mathcal E(\kappa)\sim \varepsilon ^{2/3}\kappa^{-5/3}
\end{equation}
for wavenumber $\kappa$ below $\kappa_{\mathrm d}$ and in the viscosity limit $\nu\to 0$. 
\end{Conjecture}
The critical wavenumber $\kappa_{\mathrm d}$ is often referred as Kolmogorov's dissipation wavenumber, which separates the inertial range and dissipation range. Kolmogorov's work was attained by using statistical tools and scaling analysis. Part (i) says that the linear term $\nu\Delta u$ dominates the nonlinear effect above $\kappa_{\mathrm d}$. 
 Part (ii) gives a quantitative description of the distribution of energy in frequency domain in the inertial range.
There is enormous experimental support for the predictions. Only recently, the first author and collaborators \cite{CD-Kol, CDK} have given the first rigorous mathematical justification of part (i) in the average sense.


Kolmogorov's phenomenological theory for hydrodynamics was derived under the assumption of homogeneity, isotropy and self-similarity on the flow. However, experiments suggest that a fully developed turbulent flow may be spatially and temporally inhomogeneous, which is termed as the intermittent nature. In a general principle, intermittency is characterized as a deviation from Kolmogorov's predictions. 

In the recent work \cite{CD-Kol}, we gave a mathematical definition of intermittency dimension $\delta$ of a flow through Bernstein's inequality. For 3D flow, $\delta$ belongs to $[0,3]$. The general formulation of both Kolmogorov's dissipation wavenumber and the energy spectrum by taking into account the intermittency effect was provided as well in \cite{CD-Kol}.  Kolmogorov's theory corresponds to the extreme intermittency regime $\delta=3$, in which turbulent eddies fill the space. 

The main purpose of this section is to derive scaling laws for energy spectrum, structure functions, and the critical dissipation wavenumber that separates the inertial range from the dissipation range, for the energy forward cascade dyadic model (\ref{sys-2}). 
Numerical simulation results in Section \ref{sec-num} are employed to confirm these scaling laws. 


Denote the average dissipation rate of the kinetic energy by $\varepsilon_{u}= \nu \left<\lambda_j^2a_j^2\right>$ and the kinetic energy spectrum by $\mathcal E_u(\lambda_j)=\left<a_j^2\right>/\lambda_j$; and similarly $\varepsilon_{b}= \mu \left<\lambda_j^2b_j^2\right>$ the average dissipation rate of the magnetic energy and  $\mathcal E_b(\lambda_j)=\left<b_j^2\right>/\lambda_j$ the magnetic energy spectrum. 
\begin{Conjecture}\label{q43}
The kinetic energy spectrum of the dyadic MHD system (\ref{sys-2}) in the kinetic inertial range separated from the kinetic dissipation range by the dissipation wavenumber 
\[\kappa_{\mathrm d}^u\sim (\nu^{-3}\varepsilon_u)^{\frac{1}{\delta+1}}\]
exhibits the scaling 
\[\mathcal E_u(k)\sim \varepsilon_u^{\frac23}k^{\frac{\delta-8}{3}};\]
while the magnetic energy spectrum satisfies
\begin{equation}\notag
\mathcal E_b(k)\sim
\varepsilon_b^{\frac23}k^{\frac{\delta-8}{3}}
\end{equation}
in the ion-inertial range below the magnetic dissipation wavenumber 
\[\kappa_{\mathrm d}^b\sim \left(\mu^{-3}  \varepsilon_{b} \right)^{\frac{1}{\delta+1}}.\]
\end{Conjecture}


\noindent{\textbf{Heuristic Motivation of Conjecture \ref{q43}:}}
Below is the heuristic analysis to obtain the predictions above. 
According to the $a_j$ equation of dyadic MHD model (\ref{sys-2}), 
a balance between the kinetic nonlinearity and dissipation indicates
\begin{equation} \notag
 \lambda_j^{\frac{5-\delta}{2}}a_ja_{j+1}\sim  \nu\lambda_j^2a_j,
\end{equation}
which implies 
\begin{equation}\label{scal-mhd-a}
a_{j+1}\sim \nu\lambda_j^{\frac{\delta-1}{2}}\sim \nu\lambda_{j+1}^{\frac{\delta-1}{2}}.
\end{equation}
It then follows from (\ref{scal-mhd-a}) that the kinetic energy dissipation rate of the $j$-th shell satisfies
\[\varepsilon_{u,j}:=\nu\lambda_j^2a_j^2\sim \nu^3  \lambda_j^{\delta+1}.\]
It indicates that $\lambda_j\sim (\nu^{-3}\varepsilon_{u,j})^{1/(\delta+1)}$ needs to be satisfied in order to have comparable nonlinear kinetic effect and dissipation effect, which is the motivation of the scale of the dissipation wavenumber $\kappa_{\mathrm d}^u$. 
Moreover, it implies $\nu\sim (\varepsilon_{u,j} \lambda_j^{-\delta-1})^{\frac13}$, which along with (\ref{scal-mhd-a}) indicates that 
\[\mathcal E_{u}(\lambda_j)\sim \frac{a_j^2}{\lambda_j}\sim \nu^2\lambda_j^{\delta-2}\sim \left(\varepsilon_{u,j} \lambda_j^{-\delta-1}\right)^{\frac23}\lambda_j^{\delta-2}\sim \varepsilon_{u,j}^{\frac23} \lambda_j^{\frac{\delta-8}{3}}.\]
Thus, the scaling of $\mathcal E_u$ in the conjecture follows immediately.

On the other hand, assuming a balance between the magnetic dissipation and the coupled nonlinear term in the equation of $b_j$ of (\ref{sys-2}), we infer 
\[ \mu\lambda_j^2b_j\sim \lambda_j^{\frac{5-\delta}{2}}b_ja_{j+1}\Longrightarrow a_{j+1}\sim \mu\lambda_j^{\frac{\delta-1}{2}}.\]

While the balance between the dissipation and the magnetic nonlinearity in the $a_j$ equation of (\ref{sys-2}) yields  
\[
\nu\lambda_j^2a_j\sim \lambda_j^{\frac{5-\delta}{2}}b_jb_{j+1}
\Longrightarrow b_jb_{j+1}\sim \nu\lambda_j^{\frac{\delta-1}{2}}a_j \sim \nu\mu \lambda_j^{\delta-1}. 
\]
Providing $b_j\sim b_{j+1}$, the above result gives rise to 
\begin{equation}\label{scal-mhd-b1}
b_j^2\sim \nu\mu\lambda_j^{\delta-1}.
\end{equation}
It then follows 
\begin{equation}\label{scal-mhd-b2}
\varepsilon_{b,j}:= \mu\lambda_j^2b_j^2\sim \nu\mu^2\lambda_j^{\delta+1}.
\end{equation}
Recalling $\nu\sim \varepsilon_{u,j}^{\frac13} \lambda_j^{-\frac{\delta+1}{3}}$, we have from (\ref{scal-mhd-b2})
\begin{equation}\label{scal-mhd-b3}
\mu\sim (\nu^{-1}\varepsilon_{b,j} )^{\frac12} \lambda_j^{-\frac12(\delta+1)}\sim \varepsilon_{u,j}^{-\frac16} \varepsilon_{b,j}^{\frac12} \lambda_j^{-\frac{1}{3}\delta-\frac13}.
\end{equation}
Finally, we postulate from (\ref{scal-mhd-b1}) and (\ref{scal-mhd-b3})
\begin{equation}\label{scal-mhd-b4}
\begin{split}
\mathcal E_b(\lambda_j)\sim&\ b_j^2/\lambda_j\sim \nu\mu\lambda_j^{\delta-2}\sim \varepsilon_{u,j}^{\frac16}\varepsilon_{b,j}^{\frac12}\lambda_j^{\frac{\delta-8}{3}}
\end{split}
\end{equation}
On the other hand, it follows from (\ref{scal-mhd-b3})
\begin{equation}\label{scal-mhd-b5}
\lambda_j\sim \left(\mu^{-3} \varepsilon_{u,j}^{-\frac12} \varepsilon_{b,j}^{\frac32} \right)^{\frac{1}{1+\delta}}.
\end{equation}
In the end, the balance of kinetic nonlinearity and magnetic nonlinearity in the $a_j$ equation of (\ref{sys-2}) gives 
\[\lambda_j^{\frac{5-\delta}{2}}a_ja_{j+1}\sim \lambda_{j-1}^{\frac{5-\delta}{2}}b_{j-1}^2 \Longrightarrow a_j\sim b_j.\]
Thus it is plausible to assume $\varepsilon_{u,j}\sim \varepsilon_{b,j}$. The scaling laws of the magnetic energy spectrum $\mathcal E_b$ and magnetic dissipation wavenumber $\kappa_{\mathrm d}^b$ in Conjecture \ref{q43} are thereby inspired by (\ref{scal-mhd-b4}) and (\ref{scal-mhd-b5}) respectively.

\cbdu

\bigskip

\section{Determining wavenumber}

We plan to show that there is a determining wavenumber in the following sense, for all intermittency dimension $\delta\in[0,3]$.

\begin{Theorem}\label{thm-mhd-det}
Let $(a(t), b(t))$ and $(u(t),v(t))$ be two solutions in $\ell^2$ of system (\ref{sys-2}).
 Assume that the force $f\in H^{-1}$. There exists a wavenumber $\lambda_J(t)=\lambda^{J(t)}$ such that if
\[\lim_{t\to\infty} \sum_{j=0}^J\left[\left(a_j(t)-u_j(t)\right)^2+\left(b_j(t)-v_j(t)\right)^2\right]=0,\]
then we have
\[\lim_{t\to\infty}\left(\|a(t)-u(t)\|_{\ell^2}+\|b(t)-v(t)\|_{\ell^2}\right)=0. \]
\end{Theorem}
\pf
Define 
\begin{equation}\label{def-det}
J^*(a,b)(t)=\min\{k: \lambda_{j}^{\theta-2}|a_{j}(t)|< c_0\nu, \ \ \lambda_{j}^{\theta-2}|b_{j}(t)|< c_0\mu, \ \ \forall j>k\}
\end{equation}
with a constant $c_0$ to be determined later. 
Similarly, we can define $J^*(u,v)(t)$ for the solution $(u,v)$. Then we take 
\[J(t)=\max\{J^*(a,b)(t), J^*(u, v)(t)\}.\]
We will show that $\lambda_J(t)=\lambda^{J(t)}$ is a determining wavenumber in the sense of the statement in the theorem.
Denote 
\[w_j= a_j-u_j, \ \ \ z_j=b_j-v_j, \ \ \ j\geq 0.\]
Then $(w, z)$ satisfies
\begin{equation}\label{eq-wjzj}
\begin{split}
w_j'=& -\nu \lambda_j^2w_j-\lambda_j^\theta a_jw_{j+1}-\lambda_j^\theta w_j u_{j+1}+\lambda_{j-1}^\theta (a_{j-1}+u_{j-1})w_{j-1}\\
&-\lambda_j^\theta b_jz_{j+1}-\lambda_j^\theta z_jv_{j+1}+\lambda_{j-1}^\theta (b_{j-1}+v_{j-1}) z_{j-1},\\
z_j'=& -\mu \lambda_j^2 z_j+\lambda_j^\theta a_j z_{j+1}+\lambda_j^\theta w_jv_{j+1}-\lambda_j^\theta b_jw_{j+1}
-\lambda_j^\theta z_ju_{j+1}.
\end{split}
\end{equation}
The goal is to derive energy estimate for $\sum_{j=J+1}^\infty \left(w_j^2+z_j^2\right)$ such that one can apply Gr\"onwall's inequality to establish decay of the energy. Using (\ref{eq-wjzj}) 
we estimate the energy of high modes as follows
\begin{equation}\notag
\begin{split}
&\frac12\frac{d}{dt} \sum_{j=J+1}^\infty \left(w_j^2+z_j^2\right)\\
=&\ -\nu \sum_{j=J+1}^\infty \lambda_j^2w_j^2- \mu \sum_{j=J+1}^\infty \lambda_j^2z_j^2\\
& - \sum_{j=J+1}^\infty \lambda_j^\theta w_j^2u_{j+1}+\sum_{j=J+1}^\infty \lambda_j^\theta u_j w_jw_{j+1}-\sum_{j=J+1}^\infty \lambda_j^\theta b_j w_jz_{j+1}\\
&+\sum_{j=J+1}^\infty \lambda_j^\theta v_j z_jw_{j+1}+\sum_{j=J+1}^\infty \lambda_j^\theta a_j z_jz_{j+1}-\sum_{j=J+1}^\infty \lambda_j^\theta z_j^2 u_{j+1}.
\end{split}
\end{equation}
In view of the definition of $J$, we infer from the energy equation above that 
\begin{equation}\notag
\begin{split}
&\frac12\frac{d}{dt} \sum_{j=J+1}^\infty \left(w_j^2+z_j^2\right)\\
\leq &\ -\nu \sum_{j=J+1}^\infty \lambda_j^2w_j^2- \mu \sum_{j=J+1}^\infty \lambda_j^2z_j^2\\
&+c_0\nu \sum_{j=J+1}^\infty \lambda_j^2w_j^2+c_0\nu \sum_{j=J+1}^\infty \lambda_j^2w_jw_{j+1}
+c_0\mu \sum_{j=J+1}^\infty \lambda_j^2w_jz_{j+1}\\
&+c_0\mu \sum_{j=J+1}^\infty \lambda_j^2z_jw_{j+1}
+c_0\nu \sum_{j=J+1}^\infty \lambda_j^2z_j^2
+c_0\nu \sum_{j=J+1}^\infty \lambda_j^2z_jz_{j+1}\\
\leq &\ -\nu \sum_{j=J+1}^\infty \lambda_j^2w_j^2- \mu \sum_{j=J+1}^\infty \lambda_j^2z_j^2\\
&+2c_0(\nu+\mu) \sum_{j=J+1}^\infty \lambda_j^2w_j^2+3c_0\nu \sum_{j=J+1}^\infty \lambda_j^2z_j^2.
\end{split}
\end{equation}
Therefore, if we choose $c_0$ such that 
\[c_0<\min\left\{\frac{\nu}{4(\nu+\mu)}, \frac{\mu}{6\nu}\right \},\]
then it follows
\begin{equation}\notag
\begin{split}
\frac{d}{dt} \sum_{j=J+1}^\infty \left(w_j^2+z_j^2\right)
\leq&\ -\nu \sum_{j=J+1}^\infty \lambda_j^2w_j^2- \mu \sum_{j=J+1}^\infty \lambda_j^2z_j^2\\
\leq&\ -\min\{\nu,\mu\} \sum_{j=J+1}^\infty \lambda_j^2\left(w_j^2+z_j^2\right)\\
\leq&\ -\min\{\nu,\mu\} \lambda_{J+1}^2 \sum_{j=J+1}^\infty \left(w_j^2+z_j^2\right).
\end{split}
\end{equation}
Hence, we have
\[\lim_{t\to\infty}\sum_{j=J+1}^\infty \left(w_j^2(t)+z_j^2(t)\right) =0 \]
which combined with the assumption implies
\[\lim_{t\to\infty}\left(\|a(t)-u(t)\|_{\ell^2}+\|b(t)-v(t)\|_{\ell^2}\right)=0. \]

\cbdu

\begin{Remark}\label{rk-det1}
For the 3D Navier-Stokes equation, a determining wavenumber was defined and proved to exist in the Kolmogorov regime $\delta=3$ in \cite{CDK}. In \cite{CDK}, the wavenumber is defined as
\[
\Lambda_{u}(t):=\min\{\lambda_q:\lambda_{p}^{-1+\frac 3r}\|u_p\|_{L^r}<c_0\nu ,~\forall p>q~\text{and}~ \lambda_q^{-1}\|u_{\leq q}\|_{L^\infty}<c_0\nu,~q\in \mathbb{N} \}
\]
where $u_p$ denotes the $p$-th Littlewood-Paley projection of $u$ and $u_{\leq q}=\sum_{j\leq q}u_j$.
For general dimension $\delta$ of intermittency, a wavenumber would be defined with conditions on the high modes such as
\begin{equation}\label{rk-eq1}
\lambda_j^{-1+\frac{3-\delta}{r}}\|u_j\|_{L^r}\leq c_0\nu.
\end{equation}
Applying the Bernstein's relation with intermittency correction
\begin{equation}\label{rk-eq2}
\|u_j\|_{L^r}\sim \lambda_j^{(3-\delta)\left(\frac12-\frac{1}{r}\right)} \|u_j\|_{L^2}.
\end{equation}
Combining (\ref{rk-eq1})-(\ref{rk-eq2}) yields
\begin{equation}\label{rk-eq3}
\lambda_j^{\frac{1-\delta}{2}}\|u_j\|_{L^2}\leq c_0\nu.
\end{equation}
Recall $\theta=\frac{5-\delta}{2}$ and hence (\ref{rk-eq3}) can be written as
\begin{equation}\label{rk-eq4}
\lambda_j^{\theta-2}\|u_j\|_{L^2}\leq c_0\nu.
\end{equation}
In view of the fact $a_j=\|u_j\|_{L^2}$ and (\ref{rk-eq4}), we see the wavenumber definition (\ref{def-det}) is consistent with that of the Navier-Stokes equation (PDE). 
\end{Remark}


\begin{Remark}\label{rk-det3}
When $\theta\leq 2$ and equivalently $\delta \geq 1$,  for a Leray-Hopf solution $(a(t), b(t))$ of (\ref{sys-2}), the determining wavenumber defined in (\ref{def-det}) is finite for all $t\geq 0$. In this situation, for solutions obtained in Theorem \ref{thm-stab1} which converge to the steady state $(a^*, b^*)$, we can find an upper bound for the determining wavenumber $\lambda_J$. 
\end{Remark}




Let $b\equiv 0$, the statement of Theorem \ref{thm-mhd-det} holds for the dyadic NSE. We expect the time average of the determining wavenumber to be bounded above by the dissipation wavenumber $\kappa_{\mathrm d}$ derived in Section \ref{sec-turbulence}. This will be addressed in the next section.

\bigskip

\section{Estimates of the determining wavenumber}

\medskip

\subsection{Dyadic NSE}
\label{sec-nse-est-wave}

Recall the determining wavenumber $\lambda_{J^u}(t)=\lambda^{J^u(t)}$ for the dyadic NSE with $J^u(t)$ defined as 
\begin{equation}\label{def-det-u}
J^u(t)=\min\{k\in \mathbb N: \lambda_j^{\theta-2}|a_j(t)|<c_0\nu, \ \ \forall \ \ j>k.\}
\end{equation}

\begin{Lemma}\label{le-nse1}
If $1\leq J^u<\infty$, we have 
\begin{equation}\notag
\lambda_{J^u}(t)\leq (c_0\nu)^{-1}\lambda_{J^u}^{\theta-1}(t)|a_{J^u}(t)|.
\end{equation}
If $J^u=\infty$, then $\|a\|_{H^{\theta-2}}=\infty$.
\end{Lemma}
\pf
If $1\leq J<\infty$, the minimum on the right hand side of (\ref{def-det-u}) can be reached; hence it follows immediately from the definition (\ref{def-det-u}) that
\begin{equation}\notag
\lambda_{J^u}^{\theta-2}(t)|a_J(t)|\geq c_0\nu,
\end{equation}
which gives 
\begin{equation}\notag
\lambda_{J^u}(t)\leq (c_0\nu)^{-1}\lambda_{J^u}^{\theta-1}(t)|a_{J^u}(t)|.
\end{equation}
If $J^u=\infty$, the minimum on the right hand side of (\ref{def-det-u}) can not be reached. Thus we have 
\begin{equation}\notag
\lambda_j^{\theta-2}(t)|a_j(t)|\geq c_0\nu, \ \ \forall \ \ j\geq 0.
\end{equation}
Consequently, we obtain
\begin{equation}\notag
\|a\|_{H^{\theta-2}}^2=\sum_{j\geq 0}\lambda_{j}^{2(\theta-2)}a_j^2\geq \sum_{j\geq 0}(c_0\nu)^2=\infty.
\end{equation}
\cbdu

\medskip

\begin{Lemma}\label{le-nse2}
We have the point-wise estimate for the determining wavenumber $\lambda_{J^u}(t)$,
\begin{equation}\notag
\lambda_{J^u}(t)-\lambda_0\lesssim \nu^{-2}\|a(t)\|_{H^{\theta-\frac32}}^2.
\end{equation}
\end{Lemma}
\pf
If $J^u=\infty$, we have 
\begin{equation}\notag
\|a(t)\|_{H^{\theta-\frac32}}\geq \|a(t)\|_{H^{\theta-2}}=\infty
\end{equation}
thanks to Lemma \ref{le-nse1}. Hence the statement of the lemma holds. Otherwise, if $J^u<\infty$, it follows from Lemma \ref{le-nse1} again
\begin{equation}\notag
\lambda_{J^u}^2(t)\leq (c_0\nu)^{-2}\lambda_{J^u}^{2\theta-2}(t)|a_{J^u}(t)|^2
\end{equation}
which yields
\begin{equation}\notag
\lambda_{J^u}(t)\leq (c_0\nu)^{-2}\lambda_{J^u}^{2\theta-3}(t)|a_{J^u}(t)|^2\lesssim \nu^{-2}\|a(t)\|_{H^{\theta-\frac32}}^2.
\end{equation}

\cbdu

\medskip

As an immediate consequence of Lemma \ref{le-nse2}, we have:
\begin{Corollary}\label{cor-nse}
Let the intermittency dimension $\delta\in[0,3]$ and equivalently $\theta=\frac{5-\delta}{2}\in[1, \frac52]$.
The determining wavenumber $\lambda_{J^u}(t)$ is locally integrable for every Leray-Hopf solution $a(t)$.
\end{Corollary}

\medskip

Another important estimate regards the upper bound of the averaged determining wavenumber by Kolmogorov's dissipation wavenumber, stated as follows.

\begin{Theorem}\label{le-nse3}
Let $\delta\in[0,3]$. The time average of the determining wavenumber $\lambda_J^u(t)$ is bounded by Kolmogorov's dissipation wavenumber $\kappa_{\mathrm d}^u$, namely
\begin{equation}\notag
\left<\lambda_{J^u}(t)\right>-\lambda_0\lesssim \kappa_{\mathrm d}^u, \ \ \forall \ \ \delta\in[0,3].
\end{equation}
\end{Theorem}
\pf
Recall from Section \ref{sec-turbulence} that 
\[\kappa_{\mathrm d}^u\sim \left(\frac{\varepsilon}{\nu^3}\right)^{\frac{1}{1+\delta}} \ \ \mbox{with} \ \ \varepsilon=\nu \lambda_0^\delta\left<\|a\|_{H^1}^2\right>. \]
If $J^u=\infty$, we have  $\|a\|_{H^{\theta-2}}=\infty$ by Lemma \ref{le-nse1} and hence $\|a\|_{H^{1}}\geq\|a\|_{H^{\theta-2}}=\infty$, which implies $\kappa_{\mathrm d}^u=\infty$. Thus the statement of the theorem holds.

If $1\leq J^u<\infty$, it follows from Lemma \ref{le-nse1} that
\begin{equation}\notag
\lambda_{J^u}^{1+\delta}\leq (c_0\nu)^{-2}\lambda_{J^u}^{2\theta-3+\delta}|a_{J^u}|^2=(c_0\nu)^{-2}\lambda_{J^u}^{2}|a_{J^u}|^2.
\end{equation}
Taking time average on both sides of the inequality above we obtain
\begin{equation}\label{est-ave}
\left<\lambda_{J^u}^{1+\delta}\right>\leq (c_0\nu)^{-2}\left<\lambda_{J^u}^{2}|a_{J^u}|^2\right>
\lesssim (c_0\nu)^{-2}\left<\|a\|_{H^1}^2\right>\sim (\kappa_{\mathrm d}^u)^{1+\delta}.
\end{equation}
Denote $\left< g\right>_{>\lambda_0}=\left< 1_{g>\lambda_0} g\right>$.
We have from Jensen's inequality that
\begin{equation}\label{est-ave2}
\left<\lambda_{J^u}\right>-\lambda_0=\left(\left<\lambda_{J^u}\right>_{>\lambda_0}^{1+\delta}\right)^{\frac{1}{1+\delta}}\lesssim \left(\left<\lambda_{J^u}^{1+\delta}\right>_{>\lambda_0}\right)^{\frac{1}{1+\delta}}.
\end{equation}
Combining (\ref{est-ave}) and (\ref{est-ave2}) we get
\begin{equation}\notag
\left<\lambda_{J^u}(t)\right>-\lambda_0\lesssim \kappa_{\mathrm d}^u
\end{equation}
for arbitrary intermittency dimension $\delta\in[0,3]$.

\cbdu

\medskip

\begin{Remark}\label{rk-nse1}
When $\delta=1$, $\theta-\frac32=\frac{2-\delta}{2}=\frac12$, by Lemma \ref{le-nse2} we have
\begin{equation}\notag
\lambda_{J^u}(t)-\lambda_0\lesssim \nu^{-2}\|a(t)\|_{H^{\frac12}}^2.
\end{equation}
 Note that $H^{\frac12}$ is a critical space for the 3D NSE in view of its natural scaling. When $\delta\geq 1$ ($\theta\leq 2$) we claim that the wavenumber $\lambda_{J^u}(t)$ is finite for all time $t\geq 0$. Indeed, if $J^u=\infty$, it follows from Lemma \ref{le-nse1} that $\|a\|_{H^{\theta-2}}=\infty$ and hence $\|a\|_{l^{2}}=\infty$, which is not true for any weak solution.   
\end{Remark}

\begin{Remark}\label{rk-nse2}
For the 3D NSE, the determining wavenumber was defined and proved to exist in the extreme intermittency case  ($\delta=0$) and the case $0<\delta\leq 3$, respectively in \cite{CDK} and \cite{CD-Kol}. The definitions of the determining wavenumber are distinct in the two situations. In each case, it was shown that the time average of the determining wavenumber is bounded from above by Kolmogorov's dissipation wavenumber $\kappa_{\mathrm d}^u$. Since then, it has been an interesting question that whether one can unify the definitions of the determining wavenumber in the two cases such that the optimal upper bound by $\kappa_{\mathrm d}^u$ can still be achieved. This question remains open for the 3D NSE. However, for the dyadic NSE, a universal definition of determining wavenumber exists for all the intermittency dimension $\delta\in[0,3]$ as indicated by Theorem \ref{le-nse3}.
\end{Remark}

We point out that the estimates presented in Lemma \ref{le-nse1}, Lemma \ref{le-nse2} and Theorem \ref{le-nse3} can be analogously established for the dyadic MHD models.  In the following, we show that for the dyadic NSE, the averaged determining wavenumber $\lambda_J^u(t)$ also has an upper bound in term of the Grashof number.

\begin{Lemma}Let $\delta\in[0,3]$. The time average of the determining wavenumber $\lambda_J^u(t)$ has the following bound, 
\begin{equation}\notag
\left<\lambda_{J^u}(t)\right>-\lambda_0\lesssim \lambda_0 \Big(\frac{G^2}{\nu T\lambda^2_0}+G^2 \Big)^\frac{1}{1+\delta}, \ \ \forall \ \ \delta\in[0,3].
\end{equation}
where $G := \frac{\|f \|_{H^{-1}}}{\nu^2\lambda^{\frac{1}{2}}_0}$ is the Grashof number.
\end{Lemma}
\pf By energy estimation on ({\ref{sys-2}}) with $b_j = 0$ for all $j \geq 0$, we obtain 

\begin{equation} \label{eng-eq} 
    \frac{d}{dt} \|a\|^2_2 + 2\nu \|a\|^2_{H^1} = 2 (f(t),a(t)) = 2 \sum_{j=0}^\infty f_ja_j.
\end{equation}\\
Following the proof of Theorem \ref{thm-ex-2}, we can show that there exists an absorbing ball 
\begin{equation*}
    B := \{a \in l^2 : \|a\|_2 \leq R \}
\end{equation*}
with $R=(\nu \lambda_0)^{-1}\|f\|_{H^{-1}}$. In other word, for any Leray-Hopf solution $a(t)$, there exist $t_0 \geq 0$ such that $a(t) \in B$, $\forall t > t_0 $. We note that the Grashof number satisfies $G=\lambda_0^{\frac12} \nu^{-1}R$.

Integrating (\ref{eng-eq}) over time from $t_0$ to $t_0 + T$ yields the energy inequality 
\begin{equation*}
      0 \leq\|a(t_0+T)\|^2_2\leq \limsup_{t \to t^+_0} \|a(t)\|^2_2  -  2\nu\int_{t_0}^{t_0+T} \|a(t)\|^2_{H^1}dt 
   + 2\int_{t_0}^{t_0+T} (f(t),a(t))dt.
\end{equation*}\\
It follows that
\begin{equation}\notag
\begin{split}
0 \leq&\ \nu^2 \lambda^{-1}_0 G^2 - 2\nu \int_{t_0}^{t_0+T}\|a(t)\|^2_{H^1}dt + 2\int_{t_0}^{t_0+T} \|f(t)\|_{H^{-1}}\|a(t)\|_{H^1}dt\\
 \leq&\ \nu^2 \lambda^{-1}_0 G^2 - \nu \int_{t_0}^{t_0+T}\|a(t)\|^2_{H^1}dt++\frac{1}{\nu} \int_{t_0}^{t_0+T}\|f\|^2_{H^{-1}}dt.
\end{split}
\end{equation}
Therefore, we have
\begin{align*}
\int_{t_0}^{t_0+T}\|a(t)\|^2_{H^1}dt \leq \nu \lambda^{-1}_0 G^2+ \nu^2 T \lambda_0 G^2
\end{align*}
and 
\begin{align} \label{timeavg1}
 \big <\|a(t)\|_{H^1}\big > := \frac{1}{T}\int_{t_0}^{t_0+T}\|a(t)\|^2_{H^1}dt \leq \frac{\nu G^2}{\lambda_0 T}+ \nu^2 \lambda_0 G^2.
\end{align}\\
In view of (\ref{timeavg1}), we deduce from Theorem 7.4 that for any $\delta\in[0,3]$
\begin{align*}
    \left<\lambda_{J^u}(t)\right>-\lambda_0 &\lesssim \Big(\frac{\epsilon}{\nu^3} \Big)^{\frac{1}{1+\delta}} 
   \lesssim \Big(\frac{\lambda^\delta_0}{\nu^2} \big <\|a(t)\|_{H^1}\big > \Big)^\frac{1}{1+\delta} \\
    &\lesssim \Big [ \frac{\lambda^\delta_0}{\nu^2} \Big ( \frac{\nu G^2}{\lambda_0 T}+ \nu^2 \lambda_0 G^2 \Big )\Big ]^\frac{1}{1+\delta} \\
    &\lesssim \lambda_0 \Big (\frac{G^2}{\nu T \lambda^2_0} + G^2 \Big )^\frac{1}{1+\delta}
\end{align*}
which completes the proof.
\cbdu



\bigskip

\section{Model with both forward and backward energy cascades}

The results established in previous sections are also valid for the model (\ref{sys-3}) with both forward and backward energy cascades, except that the fixed points take different form. Indeed, the stationary system of (\ref{sys-3}) can be written as 
\begin{equation}\label{stat-sys3}
\begin{split}
A_{j-1}^2-A_jA_{j+1}-B_{j-1}^2+B_jB_{j+1}=&\ \bar\nu \lambda_j^{2-\frac23\theta}A_j, \ \ j\geq 1,\\
-A_jB_{j+1}+B_jA_{j+1}
=&\ \bar\mu \lambda_j^{2-\frac23\theta}B_j, \ \ j\geq 1,\\
-A_0A_1+B_0B_1=&\ \bar\nu A_0-1,\\
-A_0B_1+B_0A_1=&\ \bar\mu B_0. 
\end{split}
\end{equation}
Without proof, we state the main results for (\ref{sys-3}) in the following.
\begin{Theorem}\label{thm-ex-sys3}
There exists a fixed point $(a^*,b^*)\in \ell^2\times \ell^2$ to (\ref{sys-3}).
Equivalently, there exists a solution $\{(A_j,B_j)\}$ to (\ref{stat-sys3}) with $(A,B)\in H^{-\frac13\theta}\times H^{-\frac13\theta}$.  
\end{Theorem}

We point out that the property of the steady state in Lemma \ref{le-positive} does not hold for the steady state of (\ref{sys-3}).

In the ideal case of $\nu=\mu=0$, the stationary system (\ref{stat-sys3}) is
\begin{equation}\label{stat-sys3-ideal}
\begin{split}
A_{j-1}^2-A_jA_{j+1}-B_{j-1}^2+B_jB_{j+1}=&\ 0, \ \ j\geq 1,\\
-A_jB_{j+1}+B_jA_{j+1} =&\ 0, \ \ j\geq 1,\\
A_0A_1+B_0B_1=&\ -1,\\
-A_0B_1+B_0A_1=&\ 0.
\end{split}
\end{equation}
\begin{Lemma}\label{le-sys3-ideal}
The solutions $\{(A_j,B_j)\}$ of (\ref{stat-sys3-ideal}) satisfy
\[A_j=A_0, \ \ B_j=B_0, \ \ \forall \ \ j\geq1,\]
\[A_0^2-B_0^2=1.\]
\end{Lemma}

\begin{Theorem}\label{thm-stab-sys3}
Solutions of system (\ref{sys-3}) with $\theta\leq 2$ and $0<f_0<\frac18 m_0^2$ converge to a fixed point in $\ell^2\times \ell^2$.
Consequently, the fixed point is unique.
\end{Theorem}

\begin{Theorem}\label{thm-stab2-sys3}
(Lyapunov Stability)
Let $(a(t),b(t))$ be a solution to (\ref{sys-3}) with initial data $(a_0, b_0)$ on $[0,T]$.  Let $(a^*, b^*)$ be the fixed point obtained in Theorem \ref{thm-ex-sys3}.
For any $\varepsilon>0$, there exists a constant $\delta>0$ such that 
\[\|a(t)-a^*\|_2^2+\|b(t)-b^*\|_2^2\leq \varepsilon, \ \ \ \forall \ \ 0\leq t\leq T,\]
provided \[\|a_0-a^*\|_2^2+\|b_0-b^*\|_2^2\leq \delta.\]
\end{Theorem}

\begin{Theorem}\label{thm-sys3-det}
Let $(a(t), b(t))$ and $(u(t),v(t))$ be two solutions in $\ell^2$ of system (\ref{sys-3}).
 Assume that the force $f\in H^{-1}$. There exists a wavenumber $\lambda_J(t)=\lambda^{J(t)}$ such that if
\[\lim_{t\to\infty} \sum_{j=0}^J\left[\left(a_j(t)-u_j(t)\right)^2+\left(b_j(t)-v_j(t)\right)^2\right]=0,\]
then we have
\[\lim_{t\to\infty}\left(\|a(t)-u(t)\|_{\ell^2}+\|b(t)-v(t)\|_{\ell^2}\right)=0. \]
\end{Theorem}

\bigskip

\section{Numerical simulations}
\label{sec-num}

In this section, we present some numerical results for the two models (\ref{sys-2}) and (\ref{sys-3}). Since both (\ref{sys-2}) and (\ref{sys-3}) are infinite ODE systems, a natural way to perform numerical simulation is to consider the cutoff of the models with the first $N$ modes for an integer $N>0$. Then the following question arises: how to choose the $(N+1)$-th mode in the $N$-th equation to close the system? After some numerical experiments, we find that setting $a_{N+1}$ and $b_{N+1}$ as a steady state leads to the predicted scalings. This choice is eligible thanks to the stability results established in Section \ref{sec-fix}. We also point out that the integer $N$ should not be too large due to the scaling size $\sim 2^N/L$ in the systems and computational limitations. Without loss of generality, we take $L=1$. More details will be provided below.

 \begin{figure}[!htb]
 \subfigure[]{\includegraphics[scale=0.42]{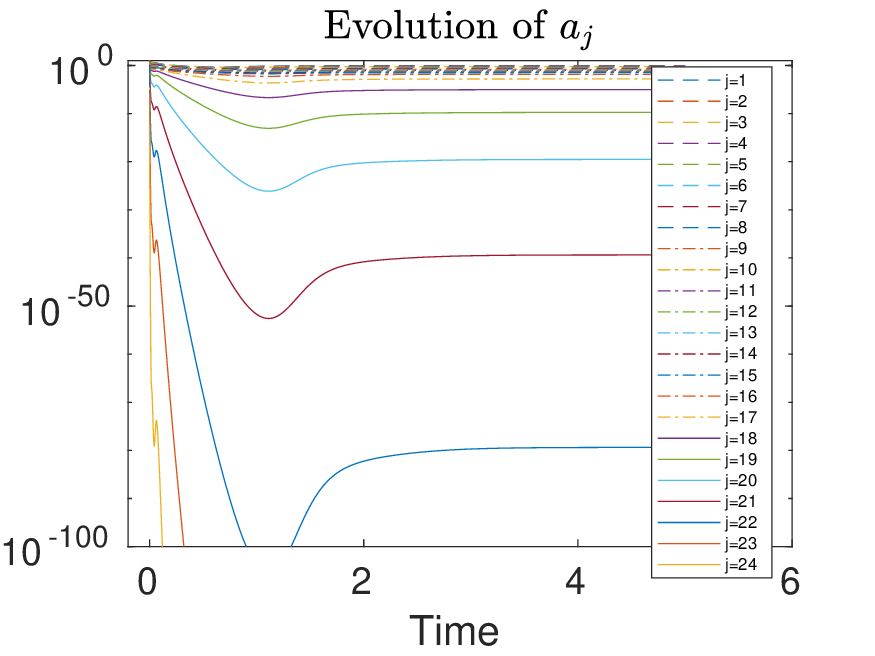} }
  \hspace{-0.2in}
    \subfigure[]{\includegraphics[scale=0.42]{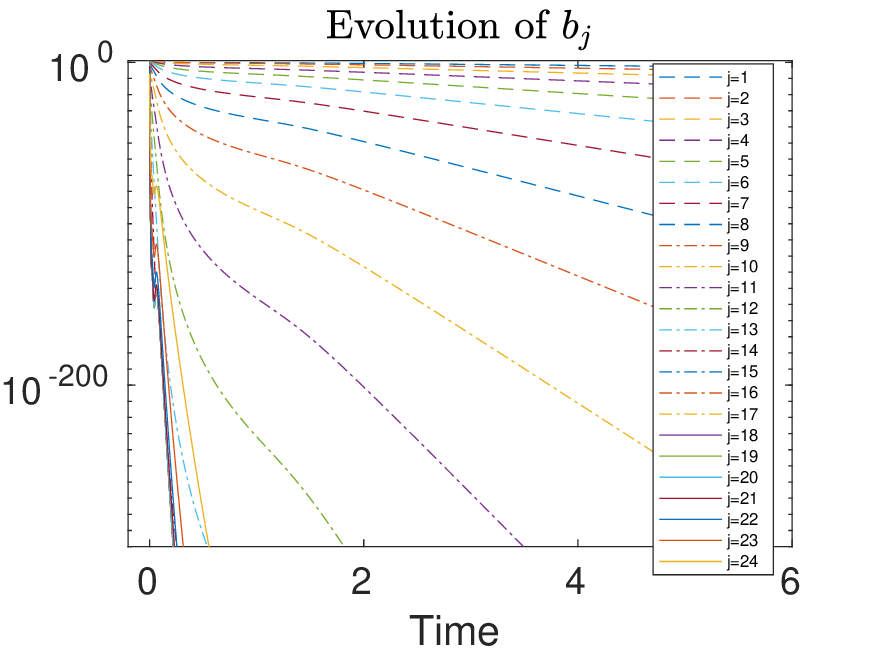} }\\
  \subfigure[]{\includegraphics[scale=0.42]{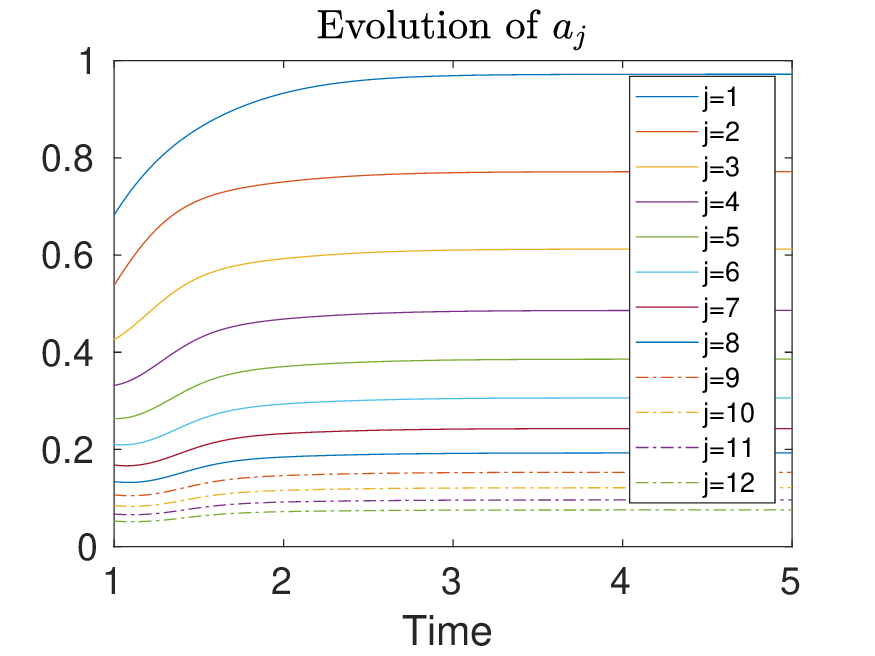} }
  \hspace{-0.2in}
    \subfigure[]{\includegraphics[scale=0.42]{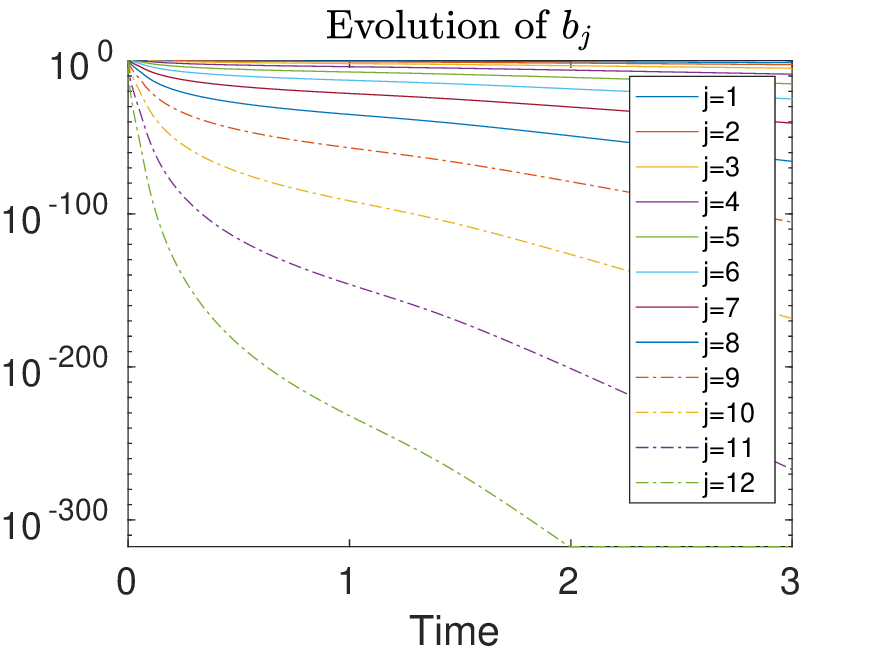} }\\
  \subfigure[]{\includegraphics[scale=0.42]{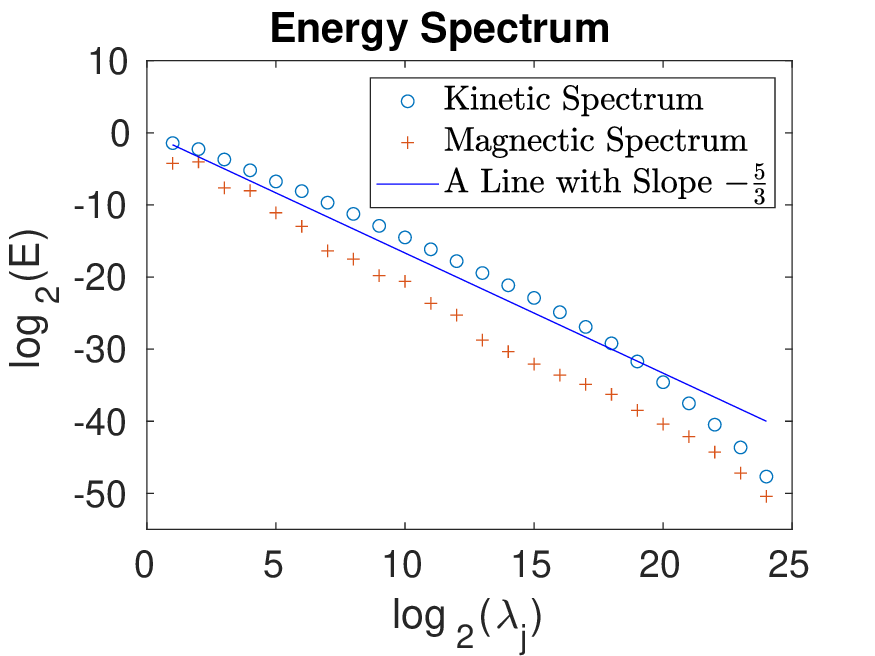} }  \hspace{-0.2in}
    \subfigure[]{\includegraphics[scale=0.42]{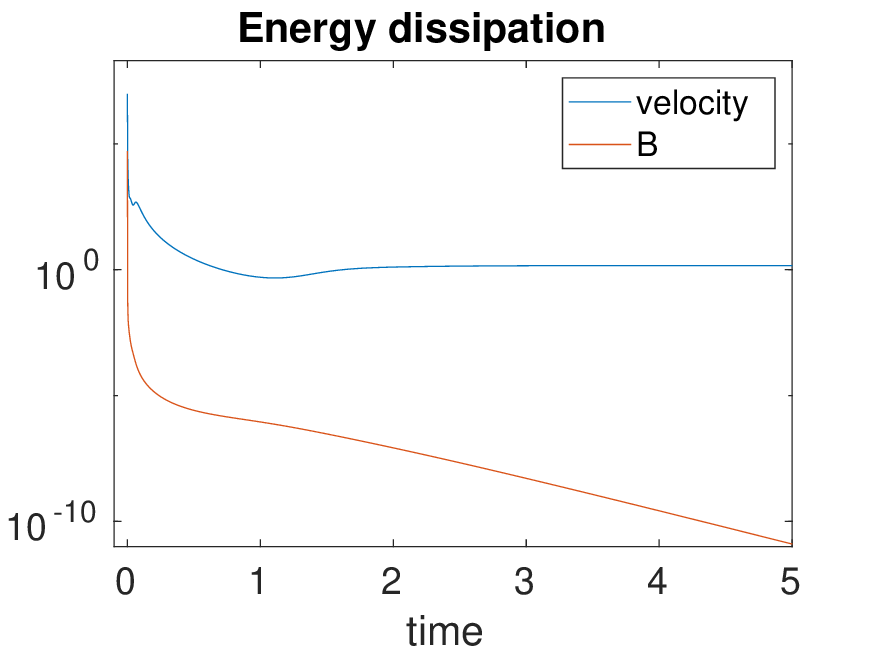} } 
  \\
  \caption{Forward cascade with $\theta=1$.} 
\label{Fig-case1}
 \end{figure}

 \medskip

\subsection{Model with forward energy cascade}
For (\ref{sys-2}), we consider the following cutoff system:
\begin{equation}\label{eq-cutoff-sys2}
\begin{split}
a'_0=& -\nu\lambda_0^2 a_0-\lambda_0^{\theta}a_0a_{1}
-\lambda_{0}^{\theta}b_0b_{1}+ f_0,\\
b'_0=& -\mu\lambda_0^2 b_0+\lambda_0^{\theta}a_0b_{1}-\lambda_{0}^{\theta}b_0a_{1} ,\\
a'_j=& -\nu\lambda_j^2 a_j-\left(\lambda_j^{\theta}a_ja_{j+1}-\lambda_{j-1}^{\theta}a_{j-1}^2\right)
-\left(\lambda_{j}^{\theta}b_jb_{j+1}-\lambda_{j-1}^{\theta}b_{j-1}^2\right), \ \ 1\leq j\leq N-1,\\
b'_j=& - \mu\lambda_j^2 b_j+\lambda_j^{\theta}a_jb_{j+1}-\lambda_{j}^{\theta}b_ja_{j+1}, \ \ 1\leq j\leq N-1.
\end{split}
\end{equation}

In the case of inviscid non-resistive dyadic MHD (\ref{sys-2}), i.e $\nu=\mu=0$, Lemma \ref{le-ex2} indicates that we have steady states $(\bar a, \bar b)$ satisfying $\bar a_j=\lambda^{\frac16\theta} f_0^{\frac12} A_0 \lambda_j^{-\frac13\theta}$ and $\bar b_j=\lambda^{\frac16\theta} f_0^{\frac12} B_0 \lambda_j^{-\frac13\theta}$ with $A_0^2+B_0^2=1$. In numerical simulation,  we can take \[a_{N+1}=\bar a_{N+1}=\lambda^{\frac16} f_0^{\frac12}A_0 \lambda_{N+1}^{-\frac13\theta},
\ \ b_{N+1}=\bar b_{N+1}=\lambda^{\frac16} f_0^{\frac12}B_0 \lambda_{N+1}^{-\frac13\theta},\]
as an approximation for sufficiently small $\nu$ and $\mu$. Therefore, we set $\nu=5\cdot 10^{-7}$ and $\mu=10^{-7}$. Fix $\lambda=2$. Other parameters are chosen as
\[\theta=1, \ \ f_0=1.5, \ \ N=24, \ \ A_0=\frac12, \ \ B_0=\frac{\sqrt 3}{2}.\]
Regarding initial data, we choose random data for $a_j(0)$ on $[0, 15]$ and for $b_j(0)$ on $[0, 4]$. Under this setting, the results of one numerical run are shown in Figure \ref{Fig-case1}.  The time evolution of $a_j$ and $b_j$ with $0\leq j\leq N$ are illustrated in Figure \ref{Fig-case1} (a) and (b), respectively. We observe that solutions $a_j$ converge to a steady state at large time and solutions remain positive; while solutions $b_j$ remain non-negative and converge to 0. To have a closer view of the trajectories, the first 12 modes of $a_j$ are plotted over time $[1,5]$ in Figure \ref{Fig-case1} (c), and the first 12 modes of $b_j$ are plotted over time $[0,3]$ in Figure \ref{Fig-case1} (d).
Figure \ref{Fig-case1} (e) shows the kinetic and magnetic energy spectrum. One can see that between mode 3 and mode 18, the kinetic spectrum lies on a straight line with slop $-\frac53$. We note that when $\theta=1$ and hence $\delta=3$, the predicted kinetic energy spectrum in Section \ref{sec-turbulence} has the scaling $\mathcal E_u(k)\sim k^{\frac{\delta-8}{3}}= k^{-\frac{5}{3}}$ in the inertial range. Therefore the numerical result justifies the scaling law. Moreover, we note in Figure \ref{Fig-case1} (e) that after mode 18, the spectrum becomes steeper, which is suggestive of the dissipation range. In fact, in view of the scaling analysis of Section \ref{sec-turbulence}, the critical dissipation wavenumber that separates the dissipation range from the inertial range has the scaling
\begin{equation}\notag
\kappa_{\mathrm d}^u\sim (\nu^{-3}\varepsilon_u)^{\frac{1}{\delta+1}} =(\nu^{-3}\bar a_0f_0)^{\frac{1}{\delta+1}} \sim 17 
\end{equation}
under the choice of the parameters. Thus, the numerical simulation is in agreement with the scaling law. Regarding the magnetic energy spectrum,  it is more or less on a line with similar slope to that of the kinetic energy spectrum with some discrepancy. Figure \ref{Fig-case1} (f) shows the kinetic and magnetic energy dissipation: $\nu\sum_{j=0}^N \lambda_j^2 a_j^2(t)$ and $\mu\sum_{j=0}^N \lambda_j^2 b_j^2(t)$. Notice that kinetic energy dissipation converges to a non-zero constant and the magnetic energy dissipation converges to zero, which shows consistency with the results of Figure \ref{Fig-case1} (a) and (b).

  \begin{figure}[!htb]
  
  \subfigure[]{\includegraphics[scale=0.42]{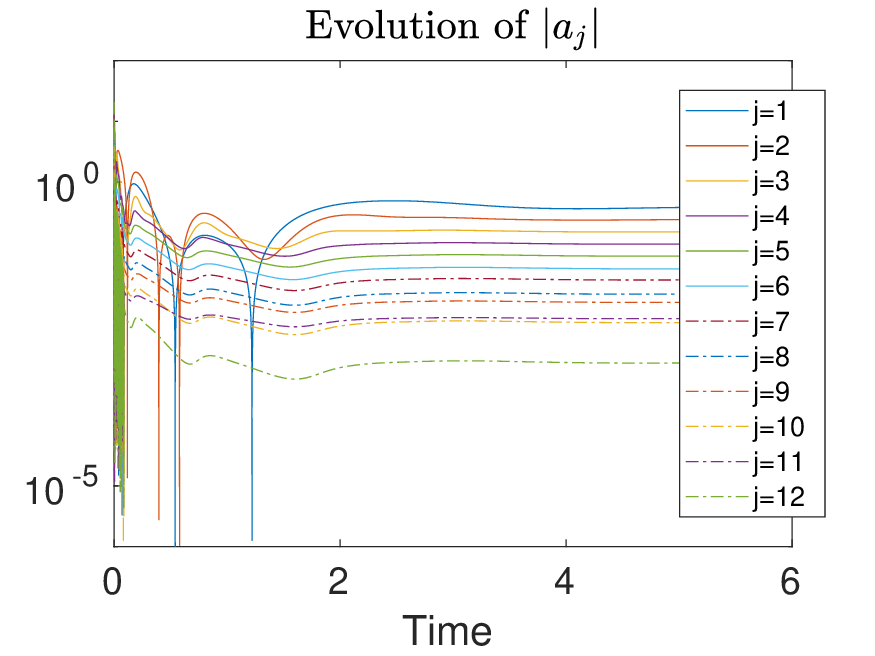} }
  \hspace{-0.2in}
    \subfigure[]{\includegraphics[scale=0.42]{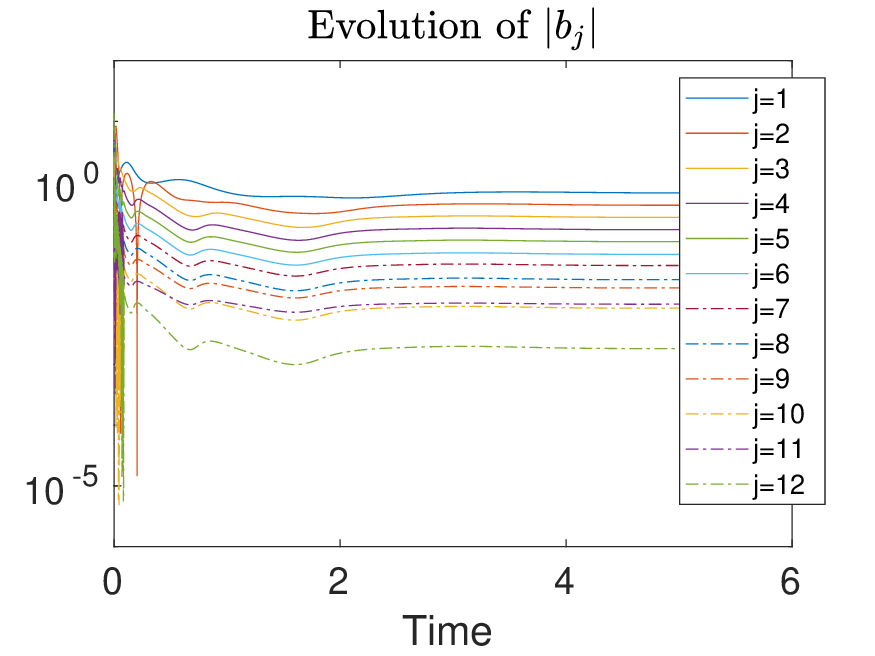} }\\
  \subfigure[]{\includegraphics[scale=0.42]{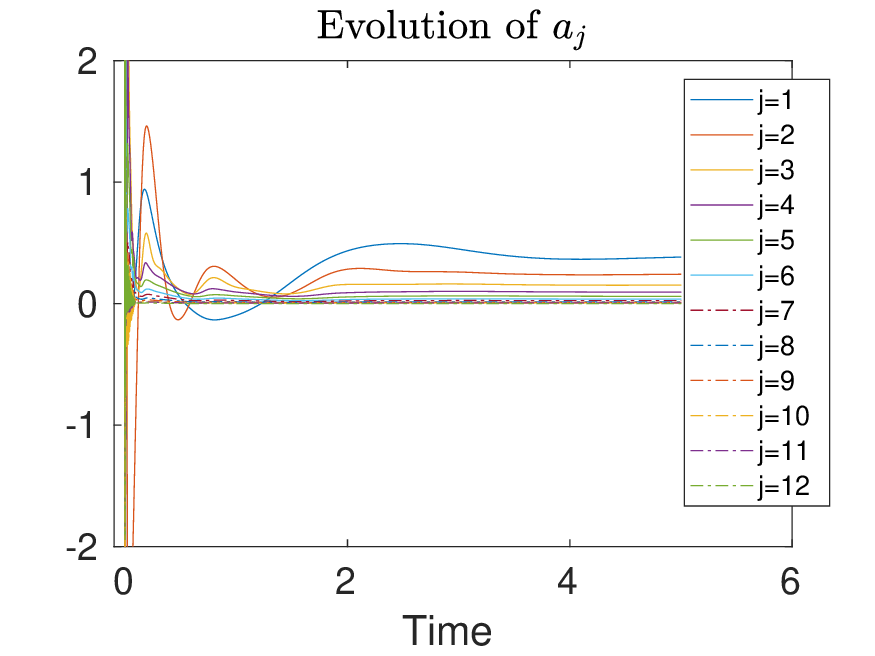} }
  \hspace{-0.2in}
    \subfigure[]{\includegraphics[scale=0.42]{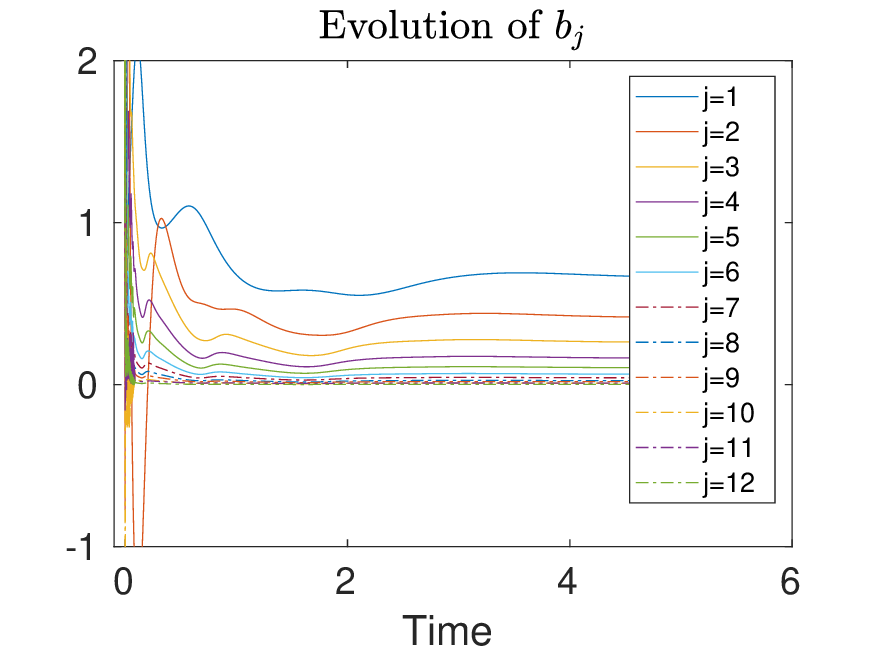} }\\
  \subfigure[]{\includegraphics[scale=0.42]{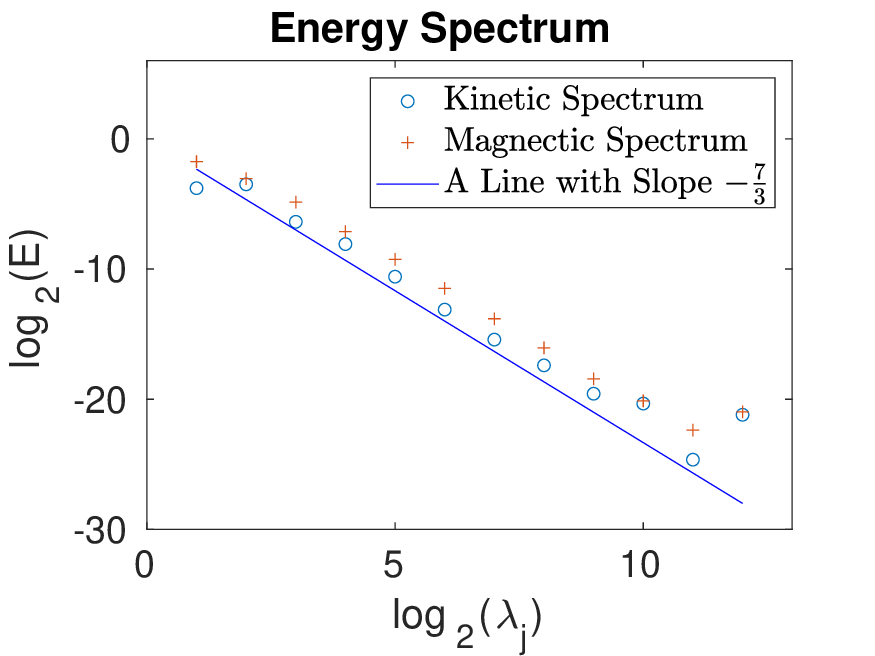} }  \hspace{-0.2in}
    \subfigure[]{\includegraphics[scale=0.42]{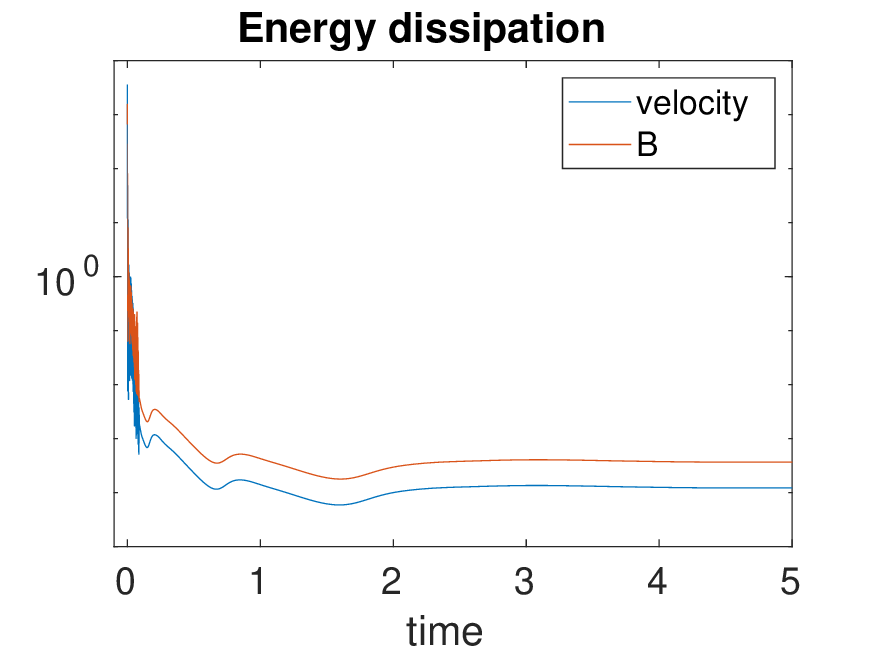} } 
  \\
  \caption{Forward cascade with $\theta=2$.} 
\label{Fig-case2}
 \end{figure}

 
 
   \begin{figure}[!htb]
  \subfigure[]{\includegraphics[scale=0.42]{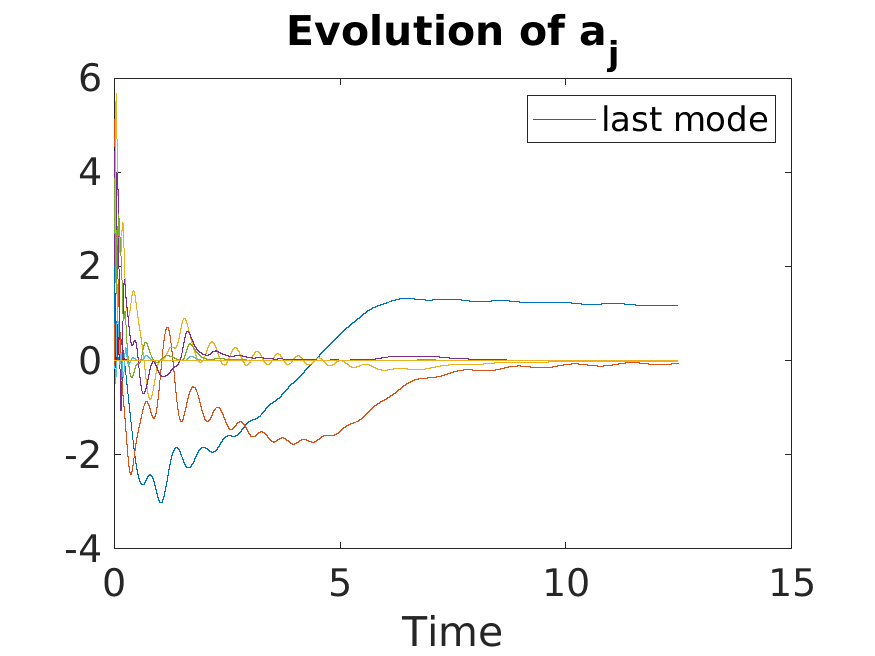} }
  \hspace{-0.2in}
    \subfigure[]{\includegraphics[scale=0.42]{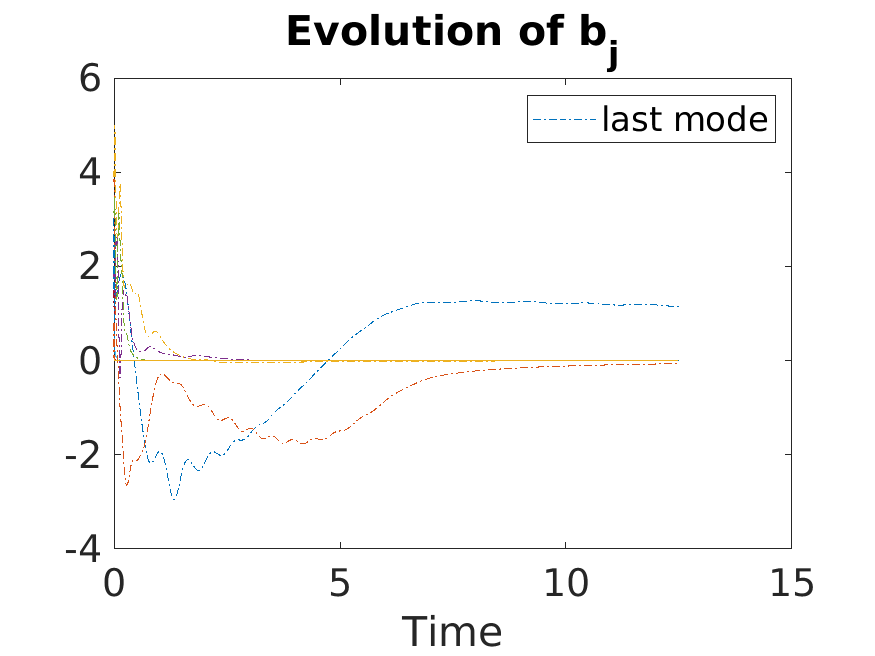} }\\
  \subfigure[]{\includegraphics[scale=0.42]{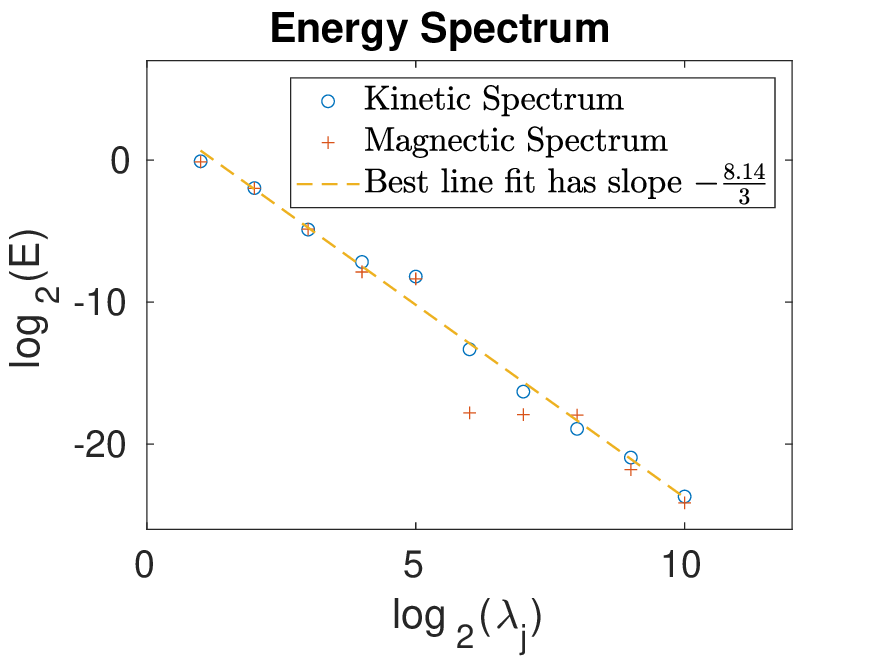} }  \hspace{-0.2in}
    \subfigure[]{\includegraphics[scale=0.42]{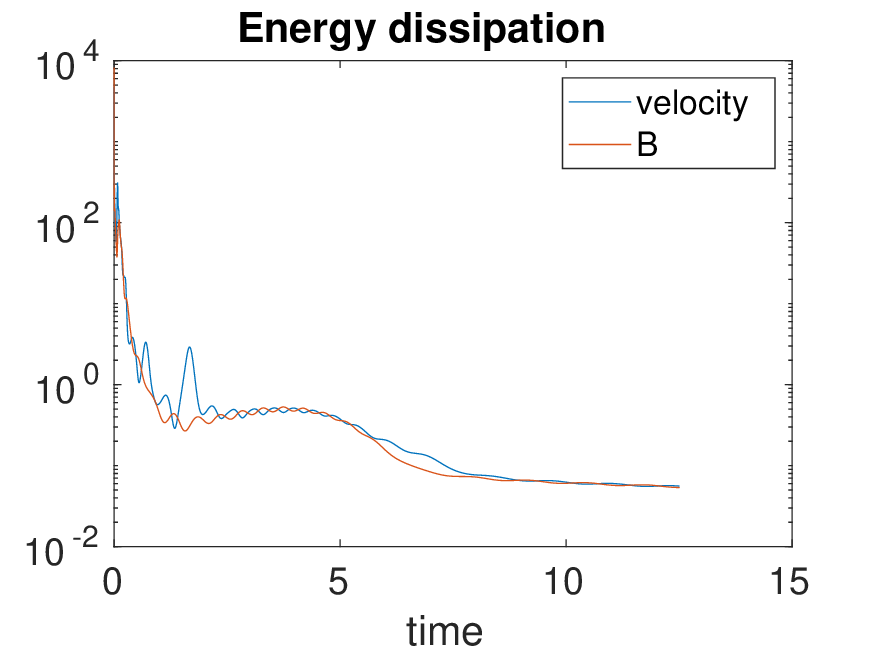} } 
  \\
  \caption{Forward and backward cascades with $\theta=1$.The slope of the best line fit for the kinetic spectrum is around $-\frac{8.14}{3}$. } 
\label{Fig-case4}
 \end{figure}

In the second setting, we take $\theta=2$ ($\delta=1$) and $N=12$, and all the other parameters remain unchanged. In this case, the intermittency dimension is smaller and the nonlinearity is stronger. The numerical simulation is shown in Figure \ref{Fig-case2}. The amplitudes of trajectories of $a_j$ and $b_j$ are plotted respectively in Figure \ref{Fig-case2} (a) and (b).
From Figure \ref{Fig-case2} (c) and (d), we observe that solutions starting from positive initial data can become negative; however, the solutions eventually converge to a positive steady state. We also note the solutions $b_j$ converge to non-zero constants. Again, the energy dissipation of Figure \ref{Fig-case2} (f) shows consistency with Figure \ref{Fig-case2} (c) and (d). Comparing Figure \ref{Fig-case2} (c), (d) and (f) and Figure \ref{Fig-case1} (a), (b) and (f), it is not hard to see that the enhanced nonlinearity influences the behaviors of the solution. The energy spectrum shown in Figure \ref{Fig-case2} (e) confirms the scaling law 
\begin{equation}\notag
\mathcal E_u(k)\sim k^{\frac{\delta-8}{3}}= k^{-\frac{7}{3}}, \ \ \mbox{with} \ \ \delta=1.
\end{equation}

\medskip

\subsection{Model with both forward and backward energy cascades}

For system (\ref{sys-3}), we consider the following cutoff approximating system:
\begin{equation}\label{eq-cutoff-sys3}
\begin{split}
a'_0=& -\nu\lambda_0^2 a_0-\lambda_0^{\theta}a_0a_{1}
+\lambda_{0}^{\theta}b_0b_{1}+ f_0,\\
b'_0=& -\mu\lambda_0^2 b_0-\left(\lambda_0^{\theta}a_0b_{1}-\lambda_{0}^{\theta}b_0a_{1} \right),\\
a'_j=& -\nu\lambda_j^2 a_j-\left(\lambda_j^{\theta}a_ja_{j+1}-\lambda_{j-1}^{\theta}a_{j-1}^2\right)
+\left(\lambda_{j}^{\theta}b_jb_{j+1}-\lambda_{j-1}^{\theta}b_{j-1}^2\right), \ \ 1\leq j\leq N-1,\\
b'_j=& - \mu\lambda_j^2 b_j-\left(\lambda_j^{\theta}a_jb_{j+1}-\lambda_{j}^{\theta}b_ja_{j+1} \right), \ \ 1\leq j\leq N-1.
\end{split}
\end{equation}

By Lemma \ref{le-sys3-ideal}, we know that the dyadic model (\ref{sys-3}) with $\mu=\nu=0$ has steady state 
$(\bar a, \bar b)$ satisfying 
\[\bar a_j=\lambda^{\frac16\theta} f_0^{\frac12} A_0 \lambda_j^{-\frac13\theta}, \ \ \bar b_j=\lambda^{\frac16\theta} f_0^{\frac12} B_0 \lambda_j^{-\frac13\theta}\]
with $A_0^2-B_0^2=1$. In numerical simulation with small $\nu$ and $\mu$,  as an approximation, we choose \[a_{N+1}=\bar a_{N+1}=\lambda^{\frac16} f_0^{\frac12}A_0 \lambda_{N+1}^{-\frac13\theta},
\ \ b_{N+1}=\bar b_{N+1}=\lambda^{\frac16} f_0^{\frac12}B_0 \lambda_{N+1}^{-\frac13\theta}.\]
Numerical experiments indicate that this model is more sensitive to the choice of the force $f_0$, viscosity $\nu$ and resistivity $\mu$. For the same size force $f_0$, $\nu$ and $\mu$ as for the approximating model (\ref{eq-cutoff-sys2}), solutions grow more rapidly. Moreover, the model seems very sensitive to the choice of initial data as well. With the same parameters, solutions for some random data grows more rapidly than other solutions. Based on these observations, we choose relatively large $\nu=\mu=0.01$ and small force $f_0=0.05$, and
\[\theta=1,  \ \ N=10, \ \ A_0=\sqrt 2, \ \ B_0=1.\]
We choose random data for $a_j(0)$ on $[0, 5]$ and for $b_j(0)$ on $[0, 4]$. The numerical results are shown in Figure \ref{Fig-case4}. The evolution of solutions $a_j$ and $b_j$ in Figure \ref{Fig-case4} (a) and (b) shows that solutions do not remain positive even though the initial data is positive. In contrast to Figure \ref{Fig-case1} (a) and (b), it takes much longer time for the solutions of (\ref{eq-cutoff-sys3}) to converge. We point out that the last modes $a_N$ and $b_N$ are far apart from other modes after convergence to a steady state; this could be an artifact due to the choice of $a_{N+1}$ and $b_{N+1}$. The energy dissipation shown in Figure \ref{Fig-case4} (d) is also consistent with Figure \ref{Fig-case4} (a) and (b). It is interesting to notice that the energy spectrum shown in Figure \ref{Fig-case4} (c) exhibits a scaling law; however, the slope is close to $-\frac{8.14}{3}$ which has a big discrepancy with the predicted slop $-\frac{5}{3}$. Many numerical runs with the same parameters show such discrepancy. Further investigation for the model with forward and backward energy cascades will be addressed in future work.




\bigskip

\section*{Acknowledgement}

M. Dai, M. Hoeller and Q. Peng are partially supported by NSF grants DMS-1815069 and DMS-2009422. X. Zhang is partially supported by NSF grants  DMS-1913120 and DMS-2208515. 
M. Dai is also grateful for the support of the Institute for Advanced Study. 

The authors would like to express their deep gratitude to the anonymous referees for their insightful suggestions which have improved the manuscript significantly.

\end{document}